\pdfoutput=1

\documentclass[preprint,12pt,3p]{elsarticle}

\usepackage{graphics}
\usepackage{amssymb}
\usepackage{amsthm}
\usepackage{amsmath}
\usepackage{float}
\usepackage{subcaption}
\usepackage{booktabs}
\usepackage{tabulary}   
\usepackage{rotating}    
\usepackage{setspace}
\usepackage{upgreek}
\usepackage{multirow}
\usepackage{pdflscape} 
\usepackage[symbol]{footmisc}

\journal{Springer}

\begin{document}

\begin{frontmatter}
\title{On the influence of alloy composition on the additive manufacturability of Ni-based superalloys}

\author[1]{Joseph N. Ghoussoub\correspondingauthor{}$^{,}$}
\author[1]{Yuanbo T. Tang}
\author[2]{William J B. Dick-Cleland}
\author[2]{Andr\'{e} A.N. N\'{e}meth}
\author[1]{Yilun Gong}
\author[1]{D. Graham McCartney}
\author[3]{Alan C.F. Cocks}
\author[1,3]{Roger C. Reed}

\address[1]{Department of Materials, University of Oxford, Parks Road, Oxford, OX1 3PH, UK}  
\address[2]{Alloyed Ltd. Yarnton, Kidlington OX5 1QU, UK}
\address[3]{Department of Engineering Science, University of Oxford, Parks Road, Oxford, OX1 3PJ, UK}

\def\correspondingauthor{\footnote{Corresponding author: joseph.ghoussoub@materials.ox.ac.uk}}

\begin{abstract}
The susceptibility of nickel-based superalloys to processing-induced crack formation during laser powder-bed additive manufacturing is studied. Twelve different alloys -- some of existing (heritage) type but also other newly-designed ones -- are considered. A strong inter-dependence of alloy composition and processability is demonstrated. Stereological procedures are developed to enable the two dominant defect types found -- solidification cracks and solid-state ductility dip cracks -- to be distinguished and quantified. Differential scanning calorimetry, creep stress relaxation tests at 1000$\,$$^\circ$C and measurements of tensile ductility at 800$\,$$^\circ$C are used to interpret the effects of alloy composition. A model for solid-state cracking is proposed, based on an incapacity to relax the thermal stress arising from constrained differential thermal contraction; its development is supported by experimental measurements using a constrained bar cooling test. A modified solidification cracking criterion is proposed based upon solidification range but including also a contribution from the stress relaxation effect. This work provides fundamental insights into the role of composition on the additive manufacturability of these materials. 
\end{abstract}


\end{frontmatter}

\newpage

\section{Introduction}

Additive manufacturing (AM) is a new process, which is particularly promising for metals and alloys \cite{herzog2016additive}. It opens up the possibility of multifunctional component design at significant geometrical precision, with close integration to CAD-based systems to enable true digital manufacturing. But like any process, it must rely upon the material feedstock for its success, and therein lies a dilemma. Can existing metallic alloys be made to process well under the extreme cooling rates and stress-states induced \cite{raghavan2016numerical, raghavan2017localized}? Indeed, much current research emphasizing process optimization assumes that this will be the case.Or alternatively, will new grades of alloy be necessary with novel compositions and behavioral characteristics designed specifically for the process \cite{tang2021alloys}? 

Here, the nickel-based superalloys are considered. They are best known for their applications in aeronautics,
hypersonics, rocketry and high temperature technologies more broadly \cite{reed2014physical}. They are an excellent test-case for the additive manufacturing process, for several reasons. First, their compositional complexity requires as many as 15 different alloying elements to be present; this broadens the freezing range and thus exacerbates solidification-related cracking whereby the remaining liquid fails to fill the partially solidified microstructure which is then pulled apart by shrinkage effects \cite{wang2017liquid}, see Figure \ref{figure1}. Recent work on solidification cracking in AM has revealed that grain boundary misorientation is critical to susceptibility; hot cracks form only on high angle grain boundaries (HAGB) \cite{chauvet2018hot} and that grain boundary segregation and ensuing liquid film stability influence crack formation \cite{kontis2019atomic, hariharan2019misorientation}. Second, the superalloys are prone to solid-state cracking due to a lack of ductility particularly in the so-called ductility dip regime of 700 to 900$\,$$^\circ$C. In the literature, the terminology used for this phenomenon is confusing, being referred to variously as ductility dip cracking, strain-age cracking and post weld heat treatment cracking. Here, we have chosen to use the all-encompassing term solid-state cracking to reflect the fact that -- in the AM process -- such cracks propagate when the material is in the solid-state \cite{boellinghaus2016cracking}. Solid-state cracks are characterized by their morphologies which are long and sharp, in contrast to the more meandering nature of solidification cracks.Work in recent years has successfully mitigated such crack formation by increased solid solution strengthening \cite{harrison2015reduction}, or by control of gamma prime ($\gamma'$) content to limit the strength induced and improve ductility \cite{murray2020defect}. And third, the multifunctional nature of their applications necessitates thin-walled and porous structures, to promote for example lightweighting and heat transfer.

The work reported in this paper was motivated with the above in mind. Here, we have avoided the temptation to fix on one alloy system, and then conduct processing trials exhaustively in an effort to work out the best combination of key process variables (KPVs) to optimise manufacturing. Instead, consistent with the ushering in of new processing methods for the superalloys  --- vacuum melting, powder metallurgy, Bridgeman single crystal processing for example --- which have always led to new compositional grades, we have taken an alternative approach: work out which alloys perform best for the process, and why. This has necessitated experimentation on a range of different alloys, including some designed specifically
for additive manufacturing.

\section{Experimental Methods}

\subsection{Choice of alloy compositions}

Twelve different superalloy compositions have been studied in this work -- five novel compositions and seven better-known so-called heritage grades, including the cast/wrought alloys IN625, IN718, and Waspaloy of moderate $\gamma^{\prime}$ content but then also IN713, IN738LC, IN939 and CM247LC which are widely employed for investment casting applications -- these display excellent creep properties on account of their high $\gamma'$ fraction. The new alloys termed ABD850AM, ABD900AM, and AM-Dev are new alloys designed specifically for the AM process, with the objective of maintaining both performance and processability. Lastly, ExpAM and ABD850AM+CB are experimental compositions, with low and high carbon and boron contents respectively, aimed at assessing the influence of C and B, as well as a very high $\gamma^{\prime}$ volume fraction in the case of ExpAM. Feedstock powders were produced via argon gas atomization; their particle size distributions each had D10 and D90 values ranging from 19.2 to 56.1$\,$$\upmu$m respectively and D50 between 32.3 and 36.5$\,$$\upmu$m. The measured alloy compositions as determined by inductively coupled plasma-optical emission spectroscopy (ICP-OES) and ICP-combustion analysis are given in Table \ref{table1}.

\begin{landscape}
	\begin{table}[H]
		\caption{Measured alloy powder compositions (wt. \%).}
		\centering
		\begin{tabular}{@{}lcccccccccccccc@{}}
			\toprule
			Alloy       & Ni  & Al  & Co   & Cr   & Fe   & Mo   & Nb   & Ta   & Ti   & W    & C     & B               & Si   & Hf  \\ \midrule
			ABD850AM    & Bal & 1.3 & 18.0 & 19.9 & -    & 2.02 & 0.64 & 0.48 & 2.25 & 4.85 & 0.010 & 0.003           & -    & -   \\
			ABD850AM+CB & Bal & 1.5 & 18.6 & 19.7 & -    & 2.03 & 0.37 & 0.57 & 2.42 & 5.05 & 0.133 & 0.007           & -    & -   \\
			ABD900AM    & Bal & 2.1 & 20.3 & 17.1 & -    & 2.09 & 1.85 & 1.21 & 2.39 & 3.06 & 0.047 & 0.003           & -    & -   \\
			AM-Dev   & Bal & 4.2 & 19.1 & 8.7  & -    & 1.20 & 3.60 & 5.50 & 0.12 & 7.10 & 0.030 & 0.005           & -    & -   \\
			CM247LC     & Bal & 5.5 & 9.3  & 8.3  & -    & 0.55 & -    & 3.20 & 0.73 & 9.66 & 0.070 & 0.020           & -    & 1.4 \\
			ExpAM       & Bal & 5.6 & 7.6  & 11.7 & -    & 2.00 & 1.10 & -    & 2.00 & 3.70 & 0.004 & \textless 0.001 & -    & -   \\
			IN625       & Bal & 0.1 & -    & 19.9 & 3.80 & 8.81 & 3.65 & -    & 0.17 & -    & 0.049 & \textless 0.001 & -    & -   \\
			IN713       & Bal & 6.0 & -    & 12.0 & 0.20 & 4.30 & 2.10 & -    & 0.70 & -    & 0.060 & 0.010           & -    & -   \\
			IN718       & Bal & 0.6 & 1.1  & 19.2 & 17.2 & 3.20 & 4.90 & -    & 0.80 & -    & 0.080 & 0.001           & 0.35 & -   \\
			IN738LC     & Bal & 3.3 & 8.5  & 15.7 & 0.02 & 1.73 & 0.87 & 1.69 & 3.29 & 2.71 & 0.110 & 0.011           & 0.03 & -   \\
			IN939       & Bal & 1.8 & 18.8 & 22.5 & -    & 0.00 & 1.00 & 1.30 & 3.60 & 1.60 & 0.160 & 0.011           & -    & -   \\
			Waspaloy    & Bal & 1.3 & 12.0 & 20.0 & 0.04 & 4.37 & 0.00 & -    & 3.00 & -    & 0.050 & 0.008           & 0.07 & -   \\ \bottomrule
		\end{tabular}
		\label{table1}
	\end{table}
\end{landscape}

\subsection{Additive manufacturing}

Processing was carried out by the laser powder bed fusion (L-PBF) method in an argon atmosphere using a Renishaw AM 400 pulsed fiber laser system of wavelength 1075$\,$nm and build plate size of 250$\times$250$\,$mm$^2$. The processing parameters employed were: laser power 200$\,$W,  laser spot size diameter 70$\,$$\upmu$m, powder layer thickness 30$\,$$\upmu$m, and pulse exposure time 60$\,$$\upmu$s. A ‘meander’ laser scan path pattern was used with hatch spacing of 70$\,$$\upmu$m and laser speed of $0.875$$\,$m/s, the path frame of reference was rotated by 67$\,$$^{\circ}$ with each layer \cite{dimter2011method}. Consistent with Renishaw's recommended practice, the speed on the borders was reduced to $0.5$$\,$m/s. To assess the composition-dependence of the severity of cracking, cubes of dimensions 10$\times$10$\times$10$\,$mm$^3$ were printed. Vertical bars of dimensions 10$\times$10$\times$52$\,$mm$^3$ were also printed to assess the mechanical properties of the alloy variants. These were manufactured with 16 inverted pyramid legs to allow for easy removal from the baseplate. Cylinders of diameter 3$\,$mm and height 1$\,$mm were produced for differential scanning calorimetry (DSC). Each alloy and geometry was processed with identical parameters, as determined to be optimal for minimizing cracking in this manufacturing system \cite{ghoussoub2020influence}.

\subsection{Assessment of additive manufacturability}

The printability was assessed by measurement of cracking severity and morphology. First, extensive quantitative stereological assessment was made with optical micrographs taken using a Zeiss Axio A1 optical microscope at 100$\times$ magnification. For this purpose, the cube samples were sectioned on (i) planes with normal corresponding to the build direction (termed XY plane) at mid-height and (ii) planes with normal perpendicular to the build direction (termed XZ). The severity of cracking was determined by imaging the XY plane at the sample height midpoint and on the XZ plane through the center; 5 images were considered per alloy. In this analysis the bulk of the sample -- 400$\,$$\upmu$m from the sample edge -- was considered. The ImageJ software was used to compare the alloys by determining the crack count density (cracks/mm$^2$) and crack length density (mm/mm$^2$) -- note that the crack length was defined as the caliper diameter corresponding to the largest line length across each crack.

In the case of a pulsed laser, solidification occurs in discrete melting events; it follows that a solidification crack must be of length equal to or smaller than a single melt pool radius. Conversely, any crack of length greater than one melt pool radius must have propagated as a solid-state crack, although of course one cannot eliminate the possibility that the solid-state crack propagated from a solidification crack. In order to provide quantitative information, the perimeter to area ratio (units $\upmu$m$^{-1}$) of cracks was assessed. Images were taken at a magnification of 100$\times$, giving a resolution high enough to identify the jagged features of solidification cracks. A further 70 images were taken of each alloy to compare the contributions of each mechanism on a representative scale; where cracking occurred, many thousands of cracks were analyzed. 

\begin{figure}[H]
	\centering
	\includegraphics[width=0.9\columnwidth]{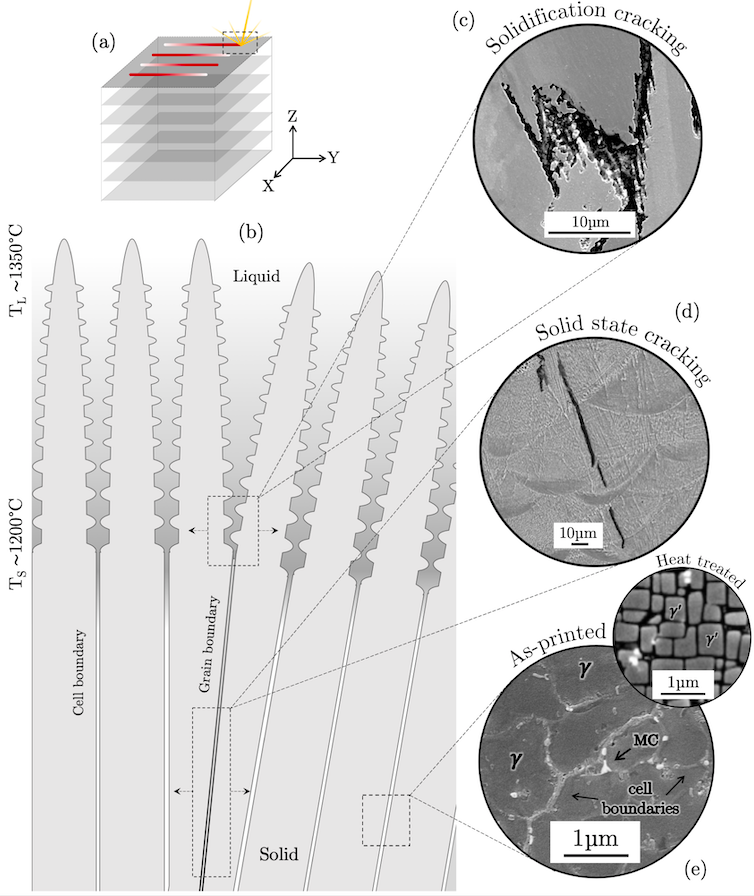}
	\caption{(a) Schematic diagram showing the L-PBF build process, whereby a laser melts powder in consecutive layers, (b) the material state and length scale at which solidification and solid-state cracking occur with corresponding SEM image examples of a (c) solidification crack and (d)  a solid-state crack in IN939 as well as (e) SEM micrographs of etched CM247LC in the as-printed condition showing the $\gamma$ matrix and MC carbide and CM247LC in the heat treated condition (reproduced from \cite{tang2021alloys}) with $\gamma^\prime$ precipitates.}
	\label{figure1}
\end{figure}

To garner further information concerning cracking mechanisms at higher spatial resolution, a Zeiss Merlin Gemini 2 field emission gun scanning electron microscope (FEG-SEM) was used equipped with an Oxford Instruments XMax 150mm/mm$^2$ energy-dispersive X-ray spectroscopy (EDX) detector. Line scans were performed to determine qualitatively the composition of carbides observed in the microstructure. Scans comprised 500 points at an accelerating voltage of 5$\,$kV and probe current 500$\,$pA resulting in an estimated interaction width and depth of 100$\,$nm. The microstructure of the alloys after manufacturing always comprised $\gamma$ and MC carbide with $\gamma'$ absent; an example of the CM247LC as-printed microstructure compared to the heat-treated microstructure following electrolytic etching with 10\% phosphoric acid at 3$\,$V direct current is given in Figure \ref{figure1}. In what follows, it is emphasized that only material in the as-printed condition is considered.

\subsection{Assessment of mechanical and stress relaxation behaviour}

\begin{figure}[H]
	\centering
	\includegraphics[width=0.6\columnwidth]{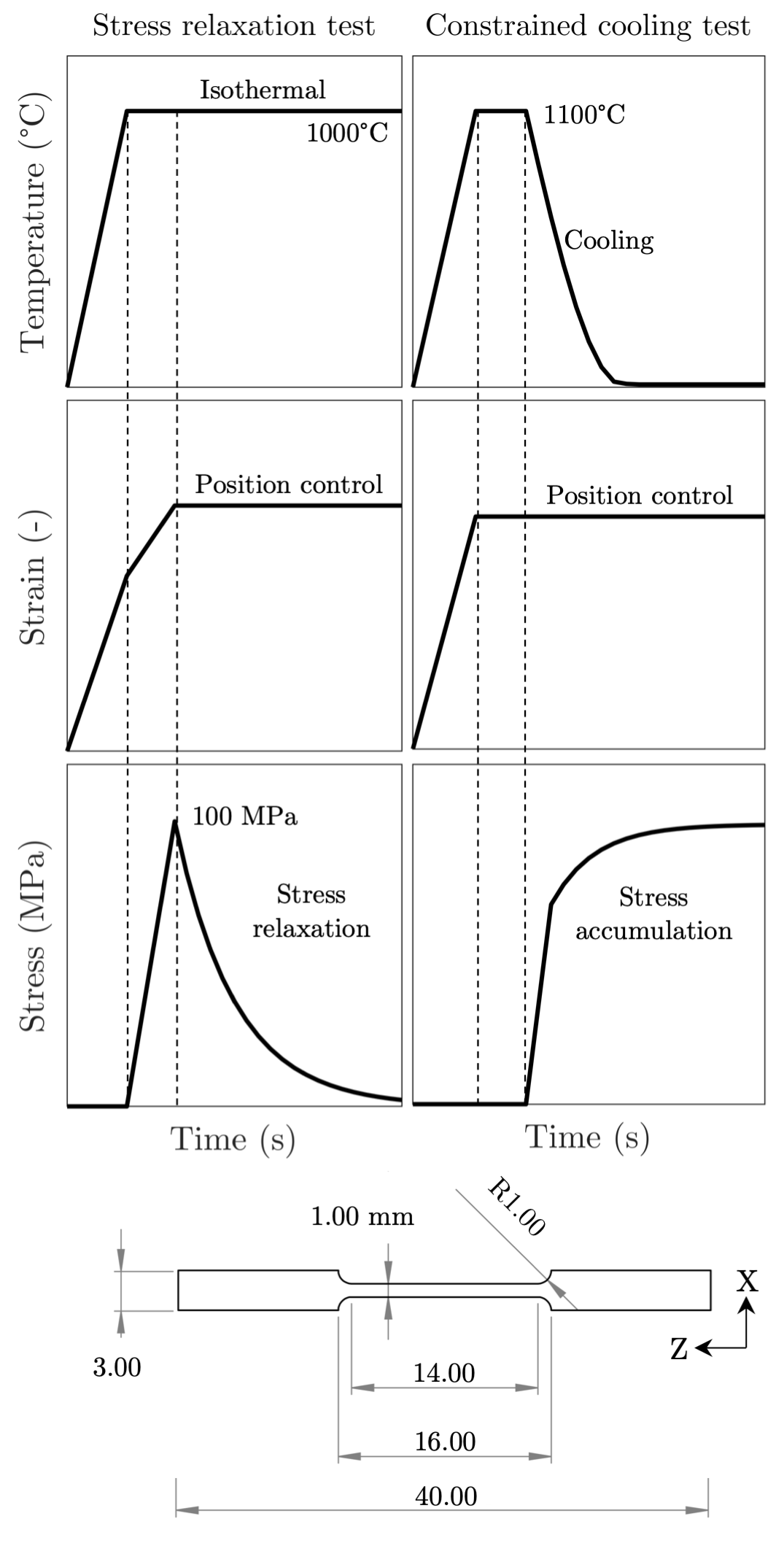}
	\caption{A schematic illustration of temperature, strain and stress profile as a function of time during the stress relaxation and constrained bar cooling test and the sample geometry used.}
	\label{figure2}
\end{figure}

\begin{figure}[H]
	\centering
	\includegraphics[width=0.6\columnwidth]{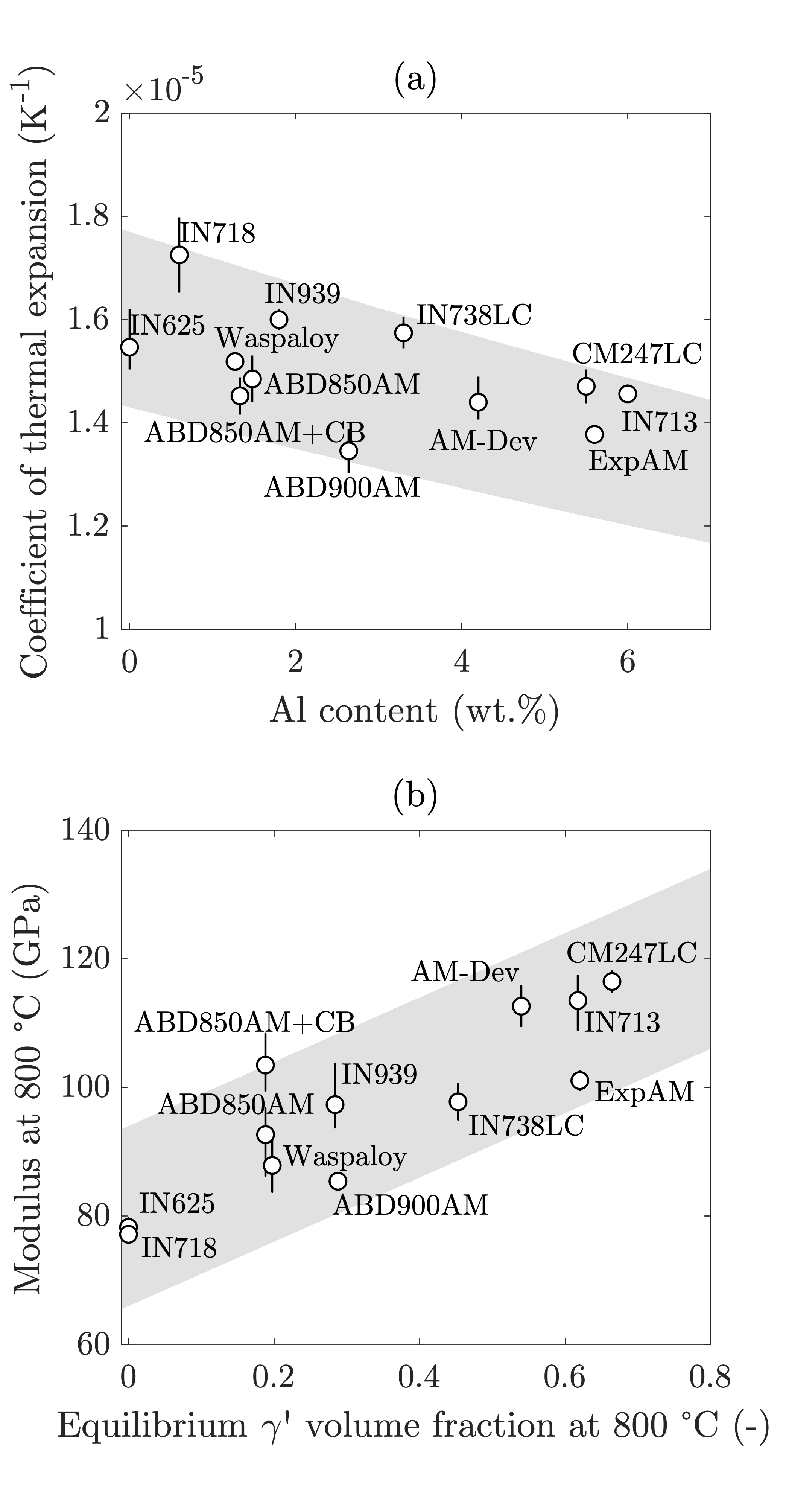}
	\caption{(a) The average coefficient of thermal expansion $\alpha$ measured during pre-test heating under free expansion conditions on the ETMT vs the Al content (wt.\%) and (b) the average elastic modulus determined at 800$\,$$^{\circ}$C vs the equilibrium  $\gamma^\prime$ volume fraction at 800$\,$$^\circ$C fraction acquired by Thermo-Calc (TTNi8 database). Error bars represent the minimum and maximum values measured.}
	\label{figure3}
\end{figure}

Uniaxial tensile, isothermal stress relaxation and constrained cooling tests were performed with an Instron electro-thermal mechanical testing (ETMT) machine. Specimens were machined with axis along the build direction using electrical discharge machining (EDM) with dimensions as given in Figure \ref{figure2}. To negate the influence of any machining-induced surface roughness, all surfaces were polished to a 4000 grit finish by hand. For each of the three types of test, joule heating under free expansion conditions ($\sigma = 0$ MPa) was employed to reach the test temperature using a heating rate of 200$\,$K/s, thus avoiding any second phase precipitation. Since it was needed for the interpretation of our results, an estimate of the average linear coefficient of thermal expansion between 300 and 800$\,$$^\circ$C was made during this pre-test sample heating as shown in Figure \ref{figure3}. The coefficient of thermal expansion is found to have a dependence on the Al content, as previously reported in \cite{morrow1975effects}.

Isothermal uniaxial tensile tests were performed for all the alloys at room temperature and at 800$\,$$^\circ$C -- this temperature being taken to be representative of where the ductility dip is likely to be severe \cite{tang2021alloys}. Specimens were strained rapidly at a rate of $10^{-2}$$\,$s$^{-1}$ to reduce the influence of any dynamic precipitation during testing. An iMetrum non-contact digital image correlation system was employed for measuring the strain; 3 repeats were performed for each alloy. The elastic modulus of each alloy was determined between 150-300$\,$MPa. The modulus determined at 800$\,$$^\circ$C correlated with increasing equilibrium $\gamma^\prime$ volume fraction, see Figure \ref{figure3}, though was likely due to the contributions of  $\gamma^\prime$ formers in solid solution of $\gamma$.

Stress relaxation tests were performed to quantify the capacity of the alloys to relieve the stress accumulated by processing. Tensile specimens were heated to 1000$\,$$^\circ$C allowing for free expansion to occur as shown in Figure \ref{figure2}, followed by a loading to 100$\,$MPa at 25$\,$MPa/s after which the sample grips were instantaneously locked into position, in order to observe the relaxation of the stress during cooling \cite{lee1971stress}. The ability to stress relax was benchmarked by comparing the stress remaining after 60$\,$s. Further stress relaxation tests at various supersolvus temperatures (1100-1200$\,$$^\circ$C) were performed on the ABD850AM+CB, Waspaloy, and  IN939 alloys by the same procedure.

A constrained bar cooling test was performed to simulate the stress build up expected of additive manufacturing. Tensile specimens were heated under free expansion conditions to 1100$\,$$^\circ$C then left for 20$\,$s to ensure thermal-mechanical equilibrium. The displacement of the ETMT grips was then fixed, and the sample was cooled to 300$\,$$^\circ$C. The accumulated stress during cooling was measured for the various alloys at cooling rates of 100$\,$K/s, 80$\,$K/s, 50$\,$K/s, and 25$\,$K/s.

\subsection{Assessment of solidification range and phase transformation temperatures}

A NETZSCH 404 F1 Pegasus DSC was used to compare solidification behaviour of the as-processed microstructure, albeit at cooling rates far lower than those experienced during AM processing. The alloys were subjected to a thermal history consisting of (i) heating to $700\,$$^\circ$C at 20$\,$$^\circ$C/min, (ii) heating from 700$\,$$^\circ$C to 1450$\,$$^\circ$C at 10$\,$$^\circ$C/min (iii) cooling to 700$\,$$^\circ$C at the same rate, and (iv) cooling to room temperature at 20$\,$$^\circ$C/min. The DSC signal from the first heat causes the transformation from a non-equilibrium additively manufactured microstructure to a homogeneous liquid. The first cool was indicative not of AM conditions, but of the solidification path from homogeneous liquid to solid at 10$\,$$^\circ$C/min. The thermophysical properties of interest derived from the DSC signal upon heating were the onset of incipient melting $T_{\mathrm{S'}}$, and the liquidus $T_{\mathrm{L}}$, as illustrated in Figure \ref{figure16}a. From cooling the onset of solidification $T_{\mathrm{L'}}$ and the solidus temperature $T_{\mathrm{S}}$ were extracted as shown in Figure \ref{figure16}b. The solidus and onset of incipient melting as well as liquidus and onset of solid nucleation temperatures were estimated following the procedures of Boettinger, Chapman, and Quested \cite{boettinger2002differential, chapman2004application, quested2009measurement}. In order to evaluate legacy crack criteria on the basis of an experimentally determined solidification path, the solid fraction as a function of temperature was estimated by considering it proportional to the ratio of the enthalpy of the transformation at $T$ to the total enthalpy of the phase transformation. This estimate is subject to possible error because other phase transformations may have occurred throughout the freezing range.

Modeling of solidification under equilibrium and Scheil-Gulliver assumptions was implemented using Thermo-Calc employing the TCNI8 and TTNi8 databases to simulate the alloy solidification path \cite{andersson2002thermo}. In the case of the Scheil models,  solidification was considered to have ended at a liquid fraction  remaining of 0.01. The propensity for backdiffusion of C to occur was deemed negligible in accordance with the dimensionless back diffusion parameter $\alpha$ \cite{kurz1989fundamentals}.

\begin{equation}
	\alpha =  \frac{4 D_{\mathrm{C}}}{\lambda_{\mathrm{cell}}^2} \frac{\Delta T_{\mathrm{freezing}}}{\mathrm{d}T/\mathrm{d}t}
	\label{eq1}
\end{equation}

\noindent where $D$ is the chemical diffusion coefficient of carbon in $\gamma$ at temperature \cite{kontis2016study}, $\lambda_{\mathrm{cell}}$ is the cell spacing, and $\Delta T_{\mathrm{freezing}}$ is the magnitude of the freezing range. Given an experimentally determined $\lambda_{\mathrm{cell}}$ of 1$\,$$\upmu$m \cite{tang2021alloys}, and typical cooling rates reported in the literature \cite{li2017efficient, masoomi2017laser, panwisawas2020additive, promoppatum2017comprehensive} the value of $\alpha$ is substantially less than unity. Hence, backdiffusion of C was deemed negligible and modeling of solidification under Scheil assumptions was considered a reasonable first approximation.

\section{Results}

\subsection{On the assessment of cracking severity, morphology, and mechanism}

\begin{figure}[H]
		\centering
		\includegraphics[width=1\columnwidth]{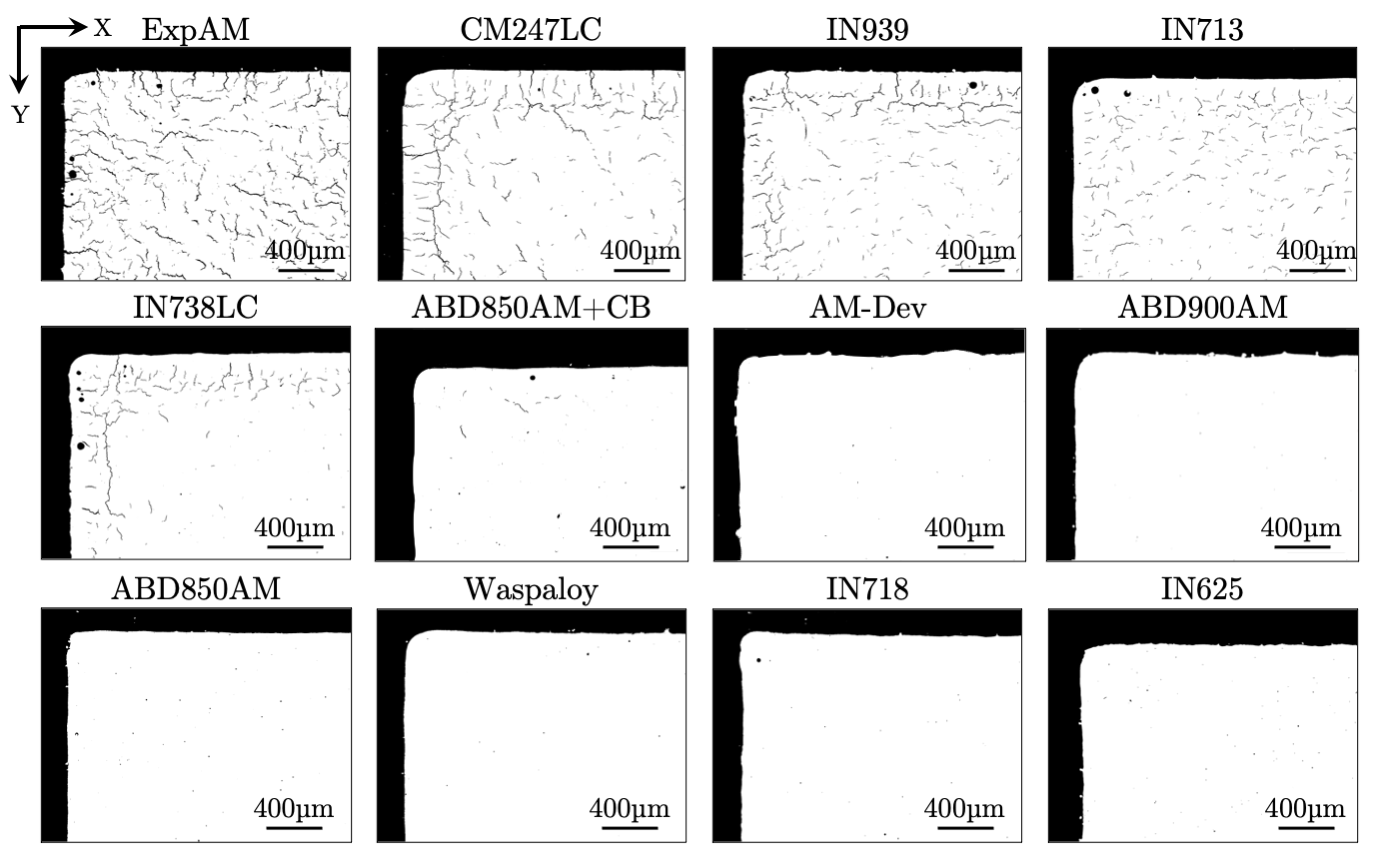}
		\caption{Optical images of the XY plane for each alloy with a binary threshold applied exemplifying cracking or lack thereof in each alloy.  It is evident the ExpAM, CM247LC, IN939, IN713, IN738LC and ABD850+CB demonstrated extensive cracking and hence are not readily processable.}
		\label{figure4}
\end{figure}

\begin{figure}[H]
		\centering
		\includegraphics[width=1\columnwidth]{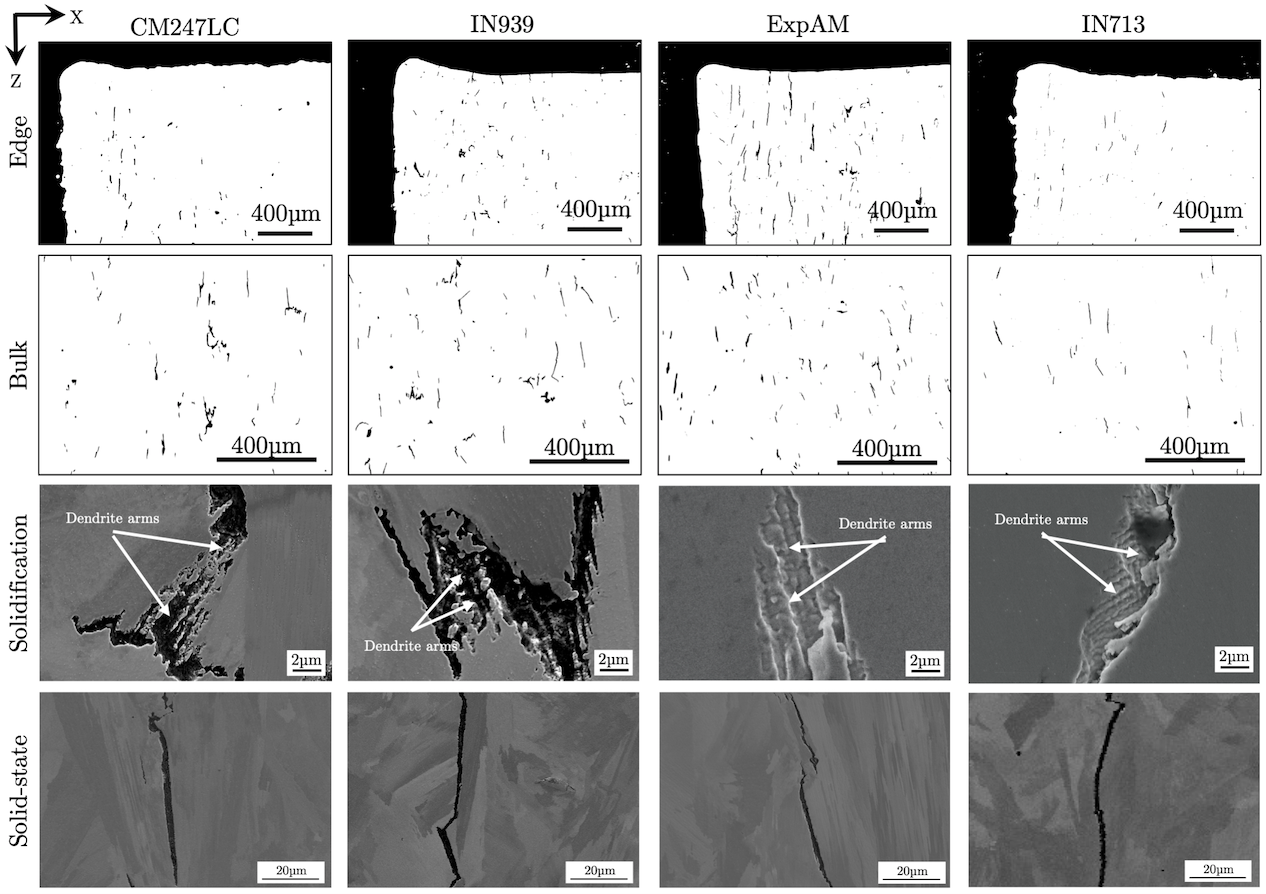}
		\caption{Optical micrographs taken of the XZ plane exemplifying cracks observed at the edge and within the bulk of CM247LC, IN939, ExpAM, and IN713, as well as SEM micrographs of solidification and solid-state cracks observed in respective alloys on the XZ plane.}
		\label{figure5}
\end{figure}

\begin{figure}[H]
	\centering
	\includegraphics[width=0.8\columnwidth]{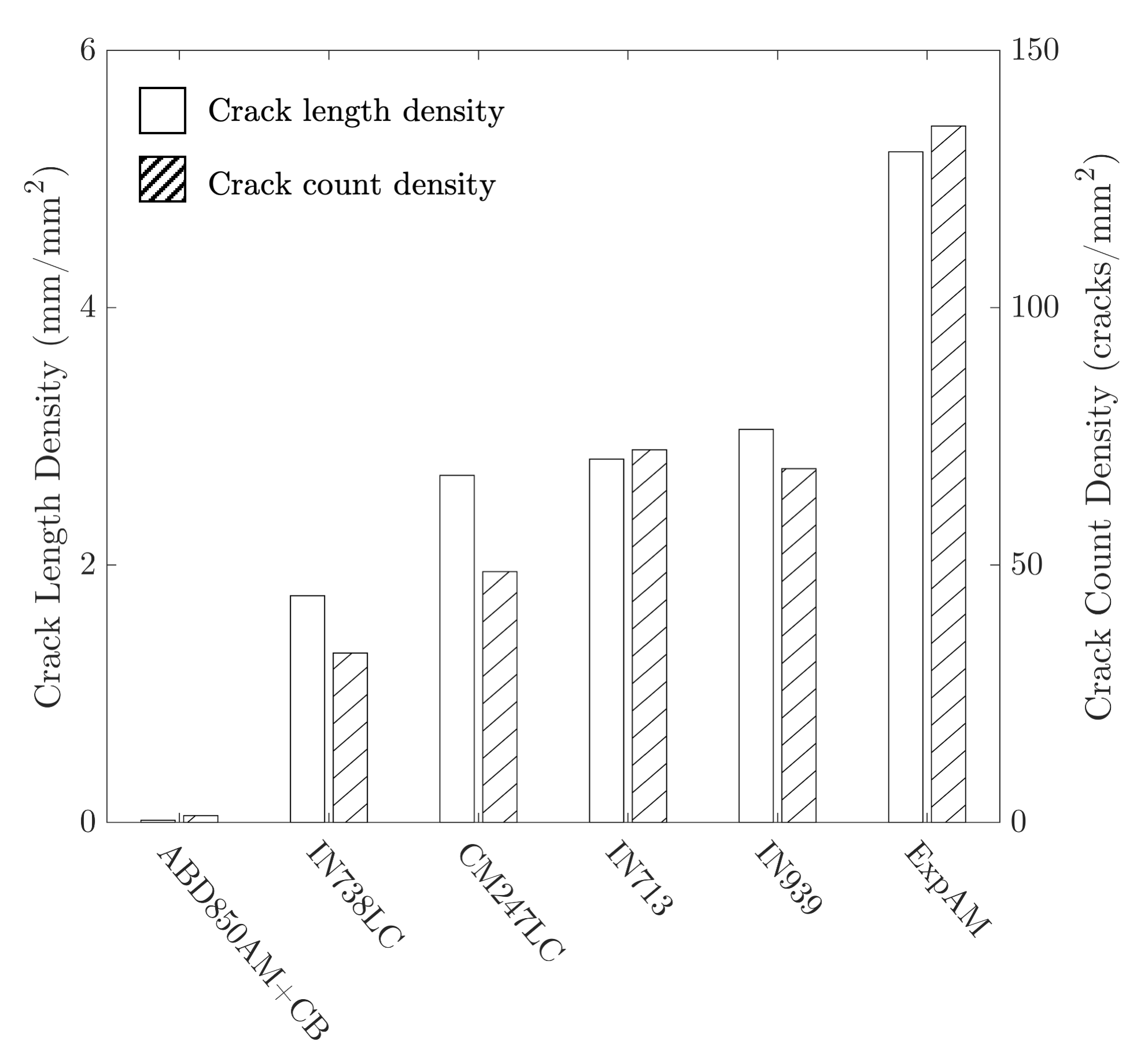}
	\caption{Bar chart showing crack length density and count count density for the six unprocessable alloys.}
	\label{figure6}
\end{figure}

\begin{figure}[H]
		\centering
		\includegraphics[width=1\columnwidth]{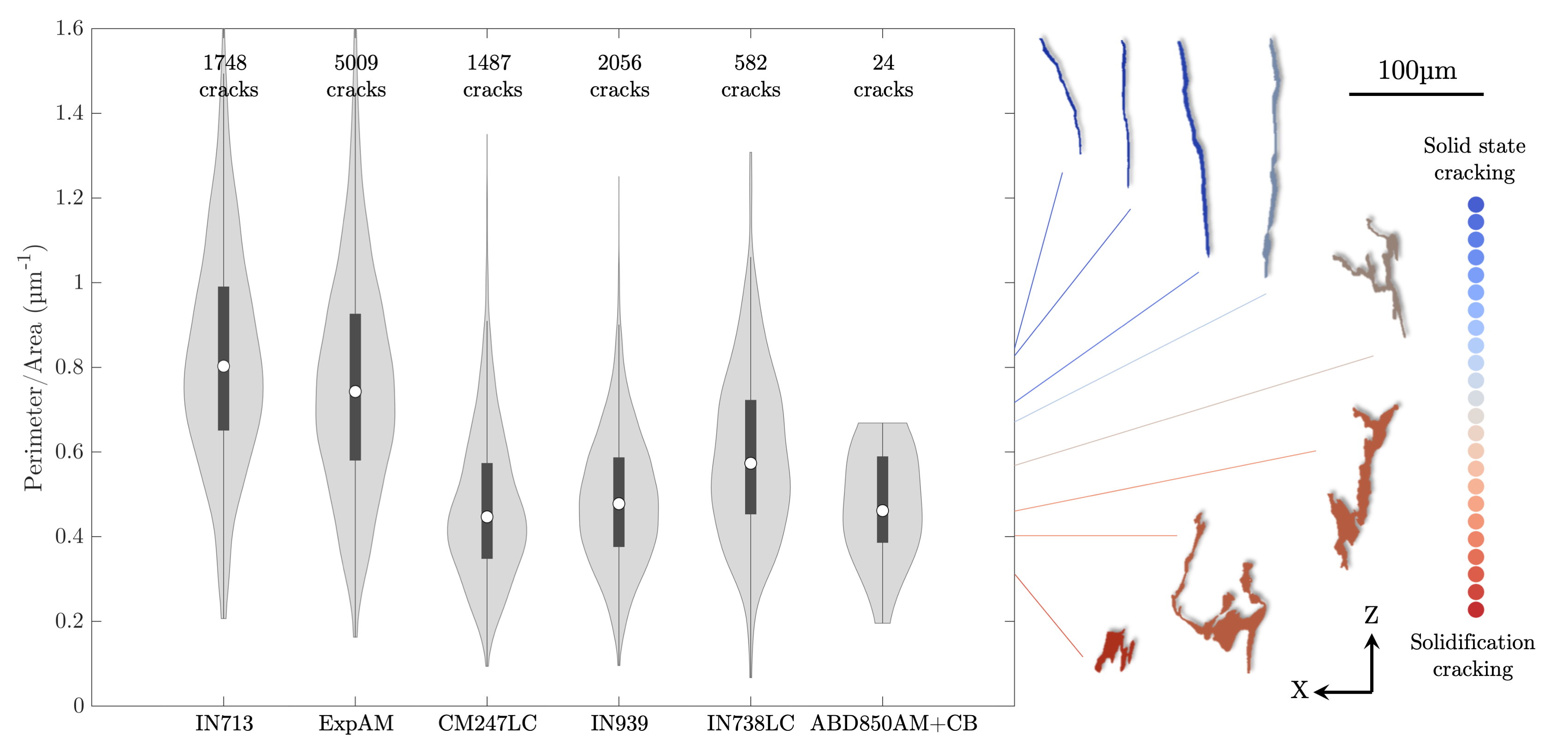}
		\caption{Violin plots showing the median, interquartile range, lower/upper adjacent values, and probability distribution of the perimeter/area ratio. As well as examples of solidification and solid-state cracks and their corresponding perimeter/area values.}
		\label{figure7}
\end{figure}

Our experimentation involving extensive characterization of cracking provides proof of the significant effect of composition on processability. This is apparent in the optical micrographs of the XY transverse section -- normal corresponding to build direction -- in Figure~\ref{figure4}. Cracking is prevalent in the legacy alloys IN713, IN738LC, IN939, and CM247LC as well as the two experimental compositions -- ExpAM and ABD850AM+CB. The other alloys examined here process largely crack-free. The distribution of cracks across the XY plane is uniform in IN713 and ExpAM, whereas the others cracked to a greater extent at the edges, where the effective laser scan speed was reduced. If one puts aside the prevalence for surface cracking for one moment and considers the cracking information for the bulk, the data for different measures of crack severity developed here -- count density and length density -- are consistent with each other, see Figure~\ref{figure6}. When examined on the longitudinal XZ plane, there is a clear disposition for the cracks to be aligned along the build direction, see Figure~\ref{figure5}, with SEM images indicating that cracks of both solid-state type and solidification type are present in all cracked alloys. Solid-state cracks tend to exhibit a long and straight morphology, whereas solidification cracks appear more jagged. 
In order to strengthen our findings further, detailed quantitative stereological analysis has been employed to quantify the extent to which solidification and solid-state cracks occur in each alloy, and to allow for distinguishing between them. We emphasize first the quantification of cracks at depth in the bulk, since these are the greatest in number and since it may be possible to improve the surface cracking susceptibility by tailoring the processing conditions at the surface still further. The morphology of cracks in different alloys is summarized by the kernel density estimate of the probability distribution of the perimeter to area ratio $\upmu$m$^{-1}$ of cracks in each of the 6 cracked alloys, see Figure \ref{figure7}. Solidification cracks, distinguished by their characteristic jagged morphology and lower perimeter to area ratio, arise to greater extent in the IN738LC, IN939, CM247LC and ABD850AM+CB. This is proven by the deviation of their perimeter to area ratio distributions, which possess a smaller median, interquartile range, and lower/upper adjacent values relative to IN713 and ExpAM. Solid-state cracking is the dominant mechanism in IN713 and ExpAM; in these alloys, cracks are longer and straighter, exemplified by the shift of their perimeter to area distributions towards greater values. Nevertheless, overlap between the distributions for some alloys is apparent for approximately 0.5 $<$ perimeter/area $<$ 0.7, suggesting that the two mechanisms are both operating simultaneously and competing against each other. Cracks in this range are of mixed mode, indicating they began as solidification cracks and then propagated further in the solid-state. The probability distribution of ABD850AM+CB is cut off at 0.19 and 0.67 as the distribution is comprised of only 24 cracks observed.

Finally, consider again the propensity for some alloys -- IN738LC, IN939, CM247LC and ABD850AM+CB -- to exhibit increased cracking at sample edges. In these alloys, cracks near edges observed on the XY plane have uniform spacing of $\approx$ 80$\,$$\upmu$m - this periodicity is attributed to the preferential nucleation of cracks to minimize strain energy \cite{tang2016numerical}. The slower scan speed used at the sample edge resulted in coarser microstructure less able to accommodate strain, with larger grains with increased texture component as compared to the bulk \cite{tang2020effect}. Furthermore, the sample edge was processed with greater energy density and larger melt pools, both of which have been observed in the literature to exacerbate solidification cracking in particular \cite{grange2020processing}. ExpAM and IN713 crack at the edges, though not comparatively more than in the bulk. However, the 4 alloys mentioned above which are prone to more solidification cracking in the bulk exhibit more cracking at the edge (relative to their bulk) as they are more sensitive to the local microstructure and processing conditions.

 \subsection{On the influence of mechanical properties on additive manufacturability}

\subsubsection{Tensile response of the alloys}

\begin{figure}[H]
		\centering
		\includegraphics[width=1\columnwidth]{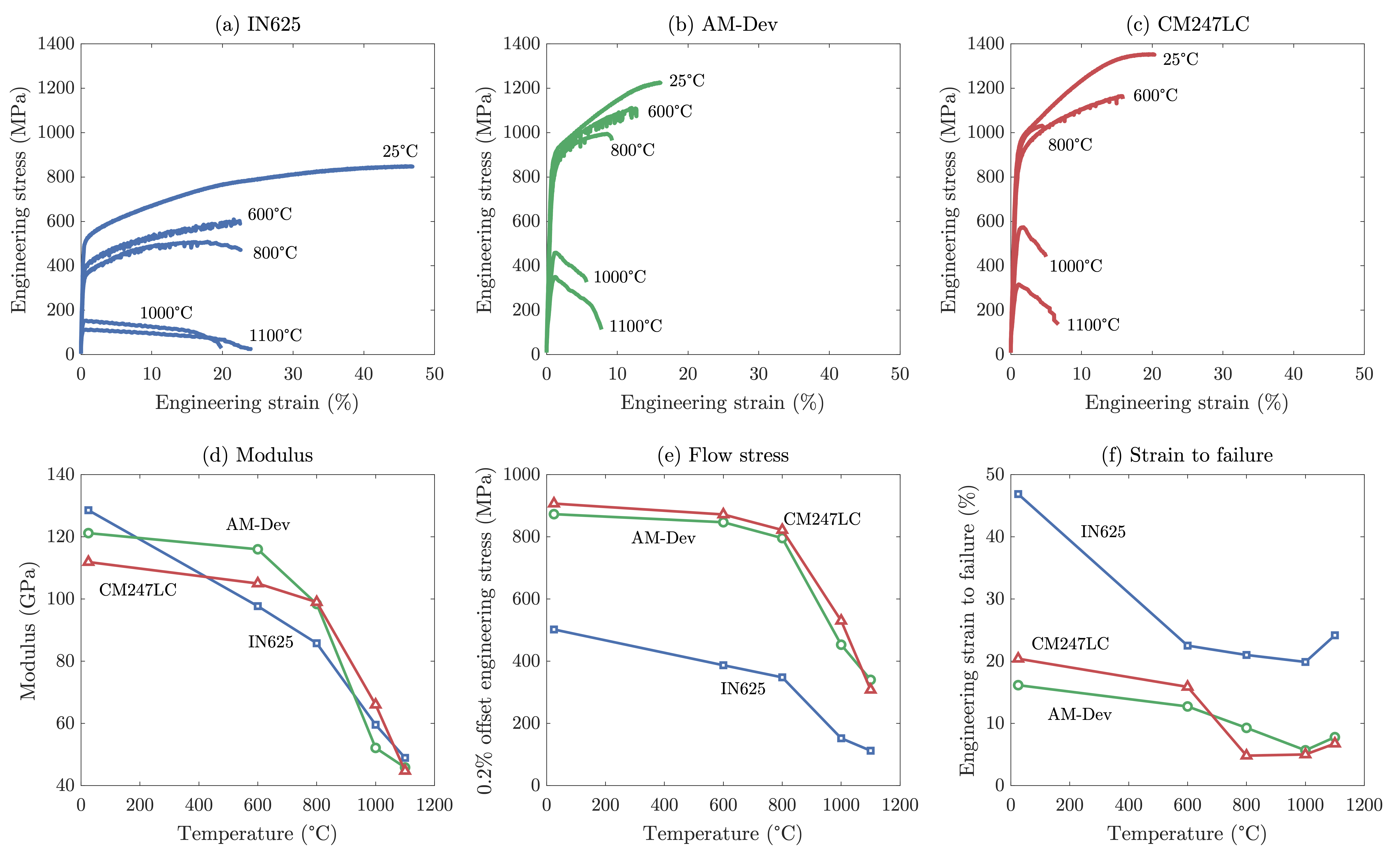}
		\caption{The isothermal tensile response of (a) IN625, (b) AM-Dev, and (c) CM247LC at 25, 600, 800, 1000, and 1100$\,$$^\circ$C in the as-printed state. Summary of the temperature dependence of (d) the elastic modulus, (e) the flow stress, and (f) the strain to failure.}
		\label{figure8}
	\end{figure}

The temperature-dependence of the tensile response at $\rm 10^{-2} s^{-1}$ is illustrated in Figure \ref{figure8}, for the alloys IN625, AM-Dev, and CM247LC. Note that the starting microstructure corresponds to the as-processed condition. The heritage alloys IN625 and CM247LC serve as lower and upper bounds respectively for flow stress across the full temperature range due to the appreciably smaller or larger extent of alloying. A significant degree of ductility is evident between 25$\,$$^\circ$C and 600$\,$$^\circ$C; however, a significant drop in ductility is found between 800$\,$$^\circ$C and 1000$\,$$^\circ$C.

\begin{figure}[H]
		\centering
		\includegraphics[width=1\columnwidth]{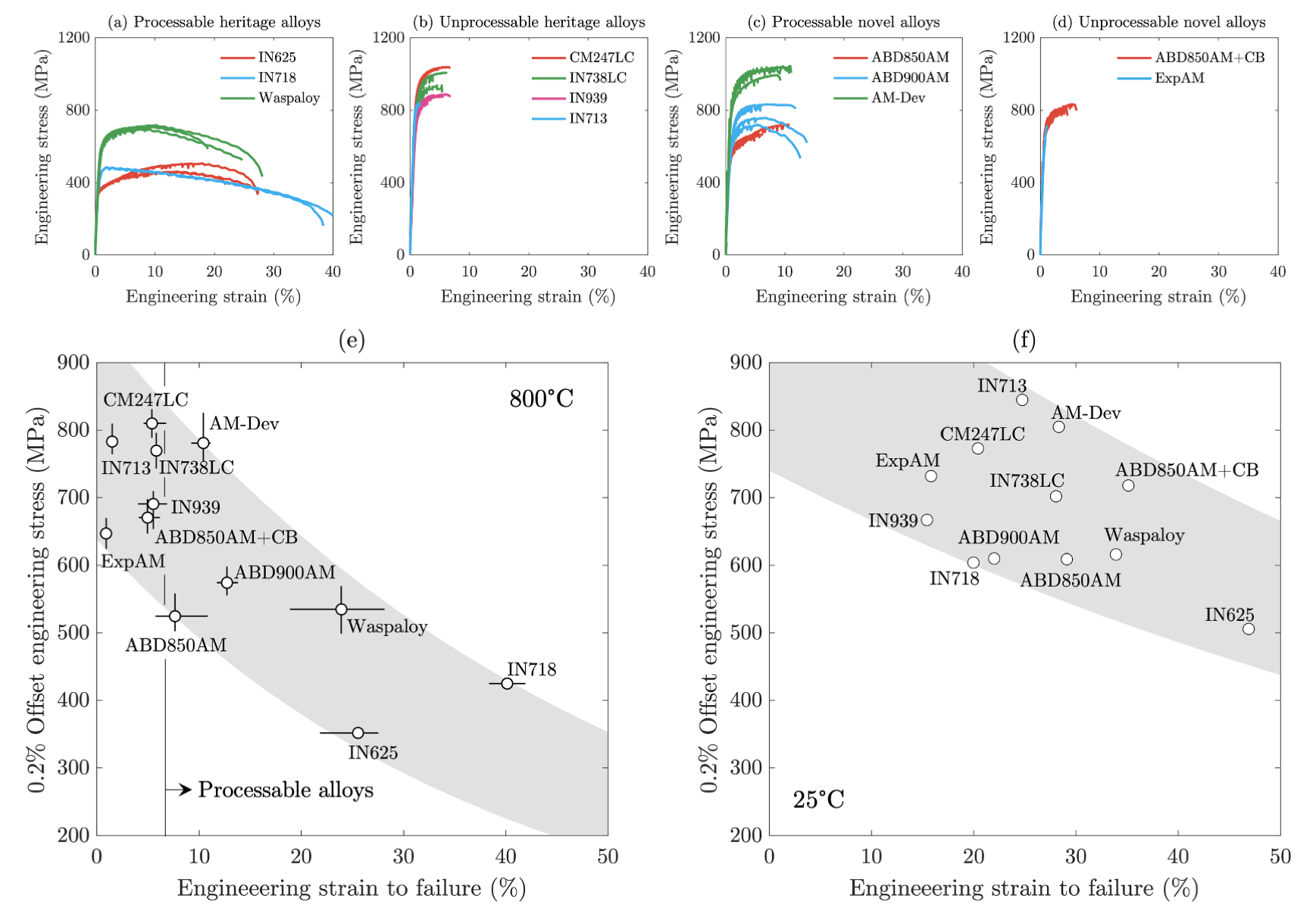}
		\caption{The tensile response of (a) the processable heritage alloys (b) unprocessable heritage alloys (c) processable novel alloys (d) unprocessable novel alloys at 800$\,$$^\circ$C. (e) the mean flow stress vs the mean engineering strain to failure at 800$\,$$^\circ$C after 3 tests, error bars show the minimum and maximum values of the 3 repeats. (f) the flow stress vs the engineering strain to failure at 25$\,$$^\circ$C.}
		\label{figure9}
	\end{figure}

The above findings have justified a thorough investigation of the composition-dependence of the strain-to-failure in the ductility-dip regime. The temperature of 800$\,$$^\circ$C was chosen for this study. The ductility at 800$\,$$^\circ$C has been found to vary within the range 1\% to 40\% for the alloys studied, see Figure~\ref{figure9}. Alloys which cracked during processing were found to exhibit significantly reduced ductility. In particular, the alloys IN713 and ExpAM -- which are prone to solid-state cracking -- failed by brittle fracture at a strain of $<$$\,$1.5\% with little necking. In this case, their engineering strain to failure has been defined as the strain at which brittle fracture takes place. For the remaining alloys, the engineering strain to failure is taken as the strain at which the sample comes apart, occurring with little/no necking, and others such as IN625 and IN718 occurring after significant necking. Serrated flow was observed in all the alloys with the exception of ExpAM, IN713, and CM247LC. This is attributed to the interaction of solute atoms and moving dislocations \cite{sharghi2008characteristics, lin2018effects}.

To summarize, it is clear that the processable alloys maintain some ductility at 800$\,$$^\circ$C whereas the unprocessable alloys do not. The engineering strain to failure and 0.2\% offset engineering stress (referred to as the flow stress) measured for each alloy at 800$\,$$^\circ$C are given in Figure~\ref{figure9} -- one can see that alloys that developed cracks during processing each fail at engineering strain to failure less than 7\% at 800$\,$$^\circ$C. Conversely, all the readily processable alloys were able to accommodate greater than 7\% engineering strain. A strength/ductility trade-off is apparent. In all 12 alloys, good ductility ($>$15\% engineering strain to failure) is observed during testing at room temperature. This confirms a negligible influence of pre-existing defects at high temperature and the relevance of a mechanism of high-temperature embrittlement.
\subsubsection{Stress relaxation behaviour}

\begin{figure}[H]
	\centering
	\includegraphics[width=1\columnwidth]{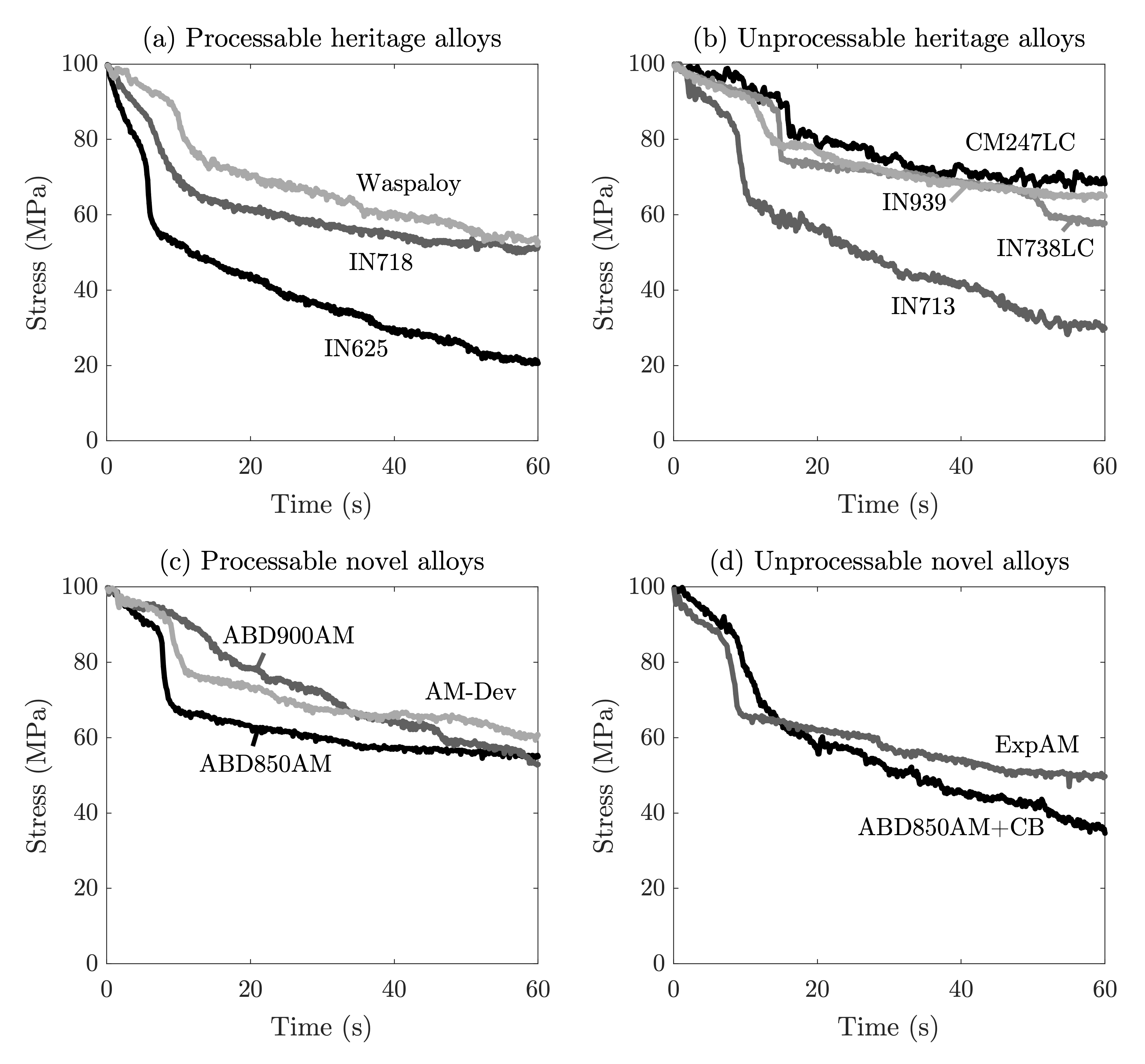}
	\caption{The stress relaxation behavior of the (a) processable heritage alloys (b) unprocessable heritage alloys (c) processable novel alloys (d) unprocessable novel alloys at 1000$\,$$^\circ$C.}
	\label{figure10}
\end{figure}

The isothermal stress relaxation responses of the different alloys at 1000$\,$$^\circ$C are compared in Figure~\ref{figure10}, for an initially imposed stress of 100$\,$MPa. The curves possess some common features -- but also differences. A significant transient is observed within the first 10$\,$s which is characterized by a rapidly decreasing stress, which develops into a more gradual decay at longer times. 

The most creep resistant alloys are those of higher Al, Ti, Ta and Nb contents such as CM247LC, IN939, IN738LC and AM-Dev. However, also apparent is that some alloys such as ABD850AM, ABD900AM and AM-Dev -- which print well -- do not relieve stress very much more slowly than those -- for example ABD850AM +CB and ExpAM -- which print poorly. It follows that the capacity for stress relaxation cannot alone be responsible for conferring additive manufacturability. 

\begin{figure}[H]
	\centering
	\includegraphics[width=1\columnwidth]{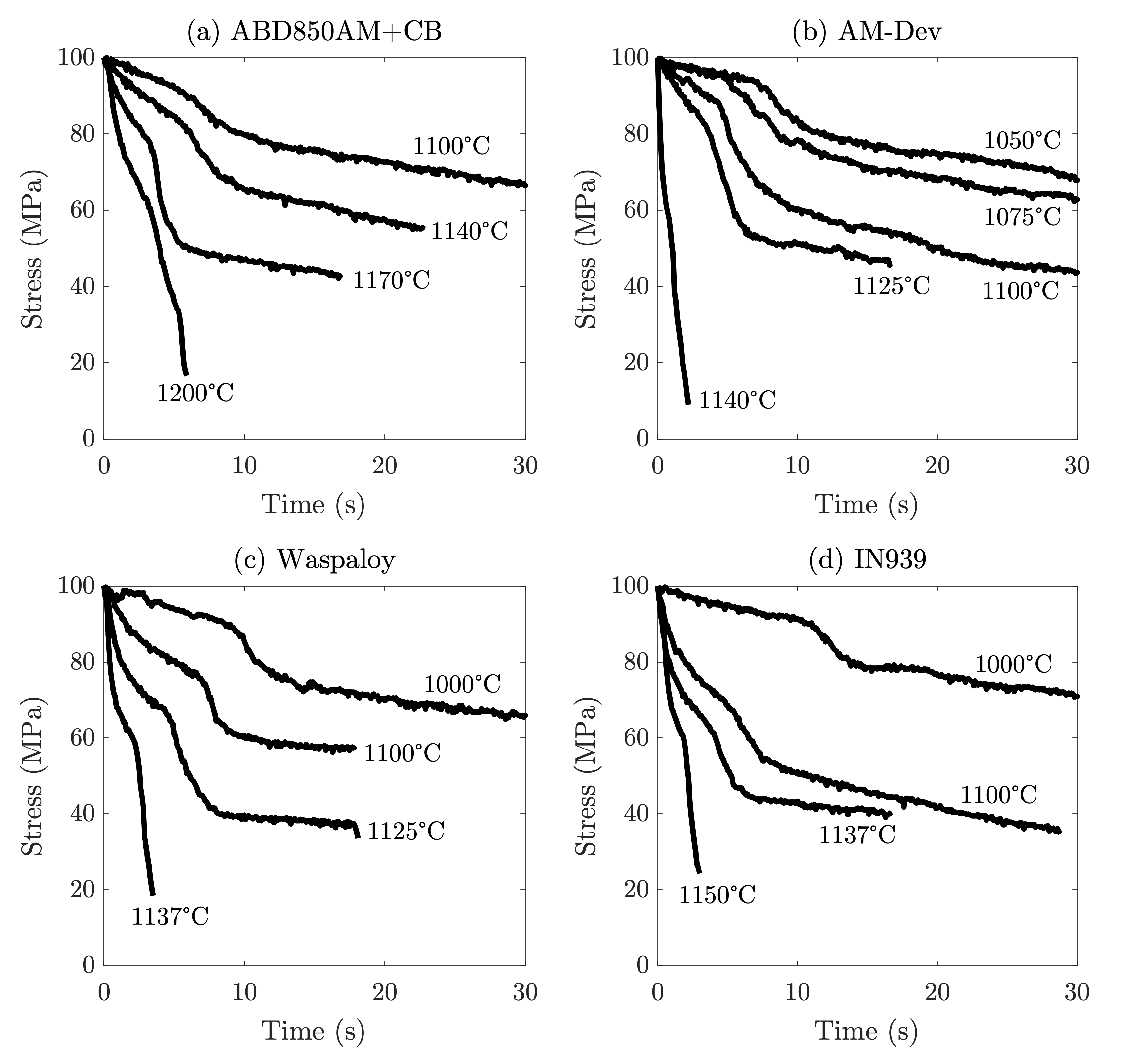}
	\caption{The stress relaxation behavior of (a) ABD850AM+CB (b) AM-Dev (c) Waspaloy and (d) IN939 at various temperatures.}
	\label{figure11}
\end{figure}

The above findings warrant further detailed quantitative analysis, and this is developed in Section~4. But the modelling presented there requires an estimate of the temperature-dependence of the stress relaxation phenomenon, and thus a first estimate of a constitutive law for this effect. For some of the alloys -- specifically IN939, Waspaloy, AM-Dev and ABD850AM+CB -- further testing has been carried out for a number of temperatures, see Figure \ref{figure11}. As expected, a strong dependence upon temperature is displayed.

\subsubsection{Stress development during constrained cooling}

\begin{figure}[H]
	\centering
	\includegraphics[width=1\columnwidth]{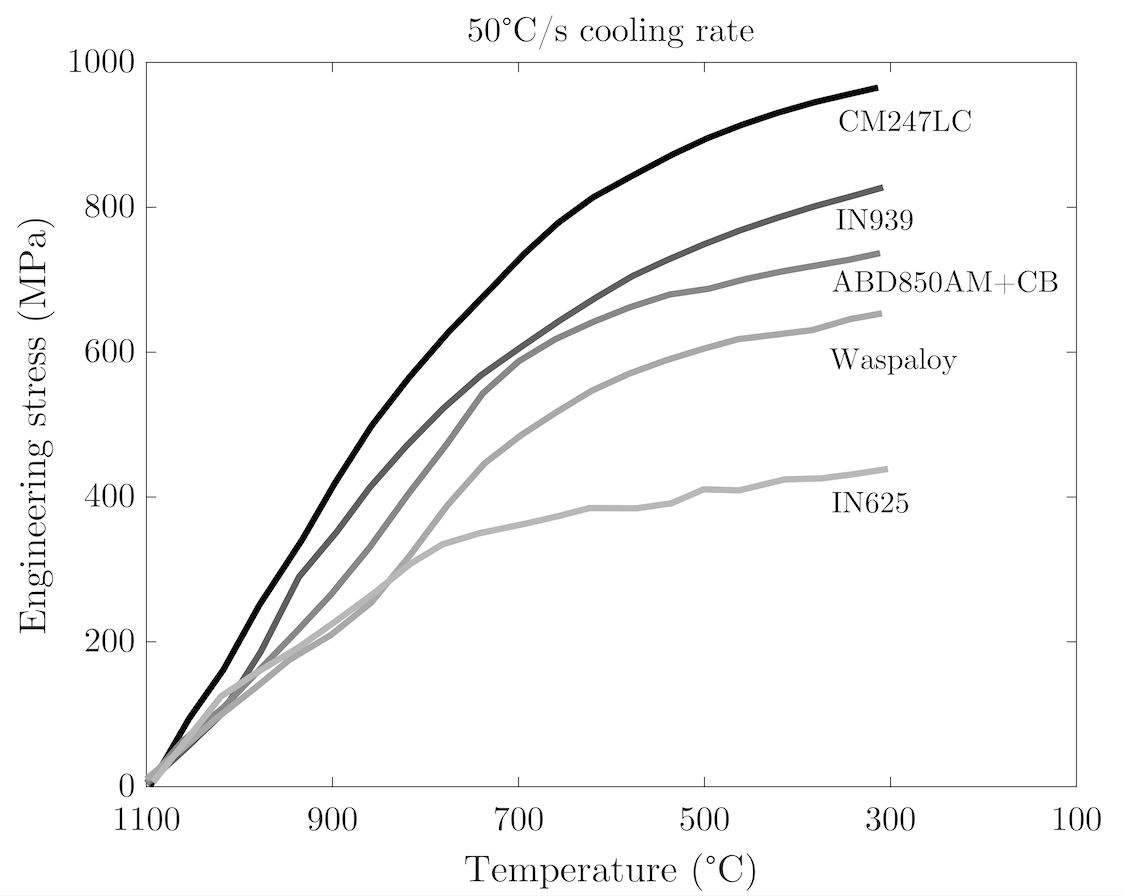}
	\caption{Comparison of the stress developed during constrained bar cooling of the CM247LC, IN939, Waspaloy, ABD850AM+CB, and IN625 alloys at a cooling rate of 50$\,$$^\circ$C/s.}
	\label{figure12}
\end{figure}

The results presented thus far indicate that lack of printability may in some way be associated with the build of stress during constrained cooling coupled with a lack of ductility in the mid-temperature regime. This motivated the assessment of the non-isothermal stress build-up during constrained cooling using the fixed-grips test involving cooling from 1100$\,$$^\circ$C at 50$\,$$^\circ$C/s, see Figure \ref{figure12}. 

At the cooling rate of 50$\,$$^\circ$C/s -- close to the fastest possible using the ETMT machine's configuration --  the alloys develop stress to different extent, see Figure~\ref{figure12}. Of the alloys assessed, stress develops in CM247LC and IN625 to the greatest and least extents respectively. Interestingly, we have never been able to replicate cracking in this constrained-bar test, at least without a further imposition of a further superimposed mechanical strain. When cooling is carried out at the slower rate of 25$\,$$^\circ$C/s the stress accumulated is concomitantly less, as discussed in the later Section~4. 

\subsection{Characterisation of as-printed microstructure and influence of freezing range}

\subsubsection{As-printed microstructure}

The as-printed microstructure in all cases comprises long columnar grains preferentially aligned in the build direction. There are three distinct characteristics: (i) cells of the matrix $\gamma$ with very small or no secondary dendrite arms; (ii) a narrow intercellular region of differing etching characteristics and therefore likely composition, and (iii) carbides residing largely in the intercellular region. It is emphasized that no $\gamma^\prime$ precipitate phase has been detected in any of the alloys in the as-printed state or following tensile testing at 800$\,$$^\circ$C. Recent data acquired via synchrotron x-ray diffraction has provided further corroboration that superalloys processed in this manner are devoid of $\gamma^\prime$ in the as-printed state \cite{tang2020effect}. The rapid cooling rate results in very fine cellular morphologies of spacing in the range 300$\,$nm to 1.5$\,$$\upmu$m depending on the proximity to the sample edge. 

\begin{figure}[H]
		\centering
		\includegraphics[width=1\columnwidth]{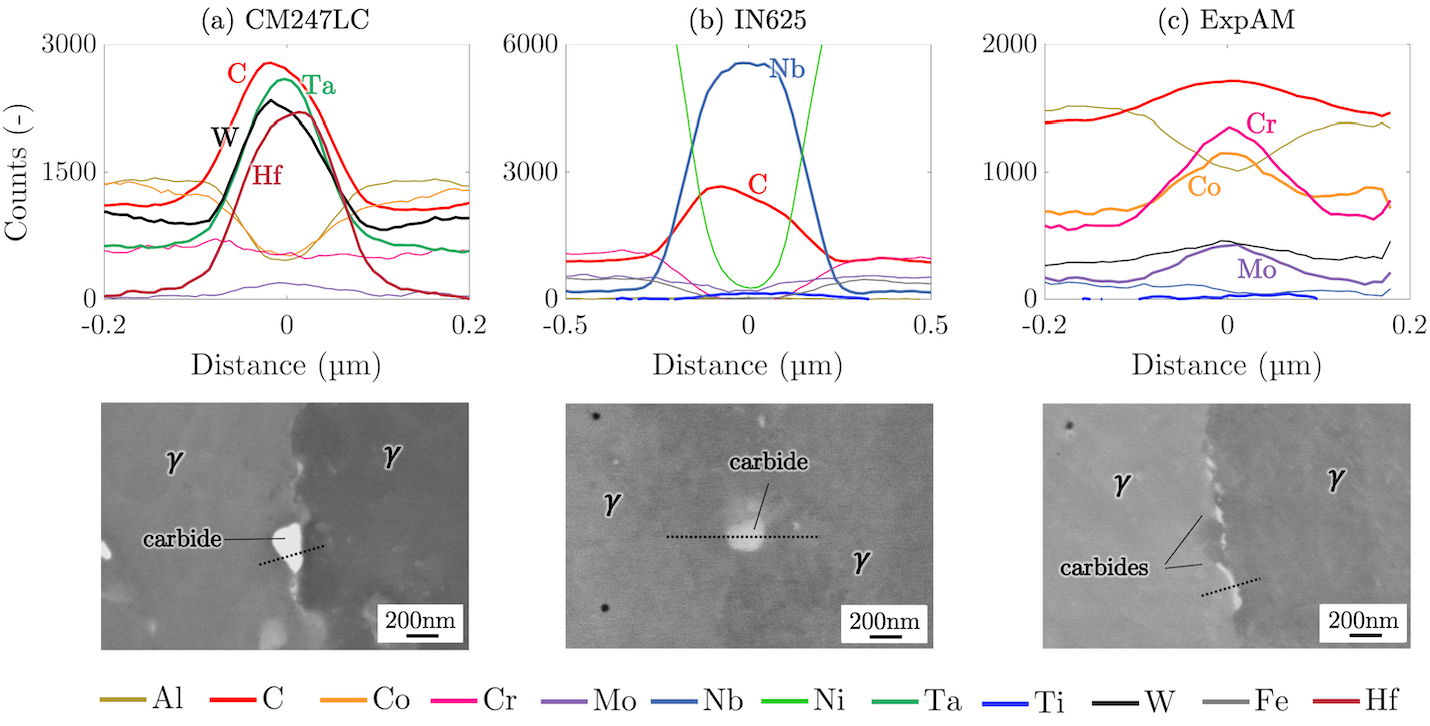}
		\caption{Backscattered electron (BSE) micrographs of the as-printed microstructure after tensile testing at 800$\,$$^\circ$C showing the $\gamma$ matrix and the intercellular carbides as well as EDX linescans across the carbides showing the carbide enrichment in (a) CM247LC, (b) IN625, and (c) ExpAM.}
		\label{figure13}
\end{figure}

Figure~\ref{figure13} shows the enrichment of the intercellular carbides in CM247LC, IN625, and ExpAM as detected by EDX linescans. A table of the enrichment observed in each alloy and confirmation from the literature is made in Table~\ref{table2}. The literature indicates that the carbides are likely of MC type of face-centered cubic structure. The presence of Cr enrichment in only ExpAM -- the most severely cracked alloy -- suggests a deleterious effect of chromium-rich carbides on processability.

\begin{figure}[H]
		\centering
		\includegraphics[width=1\columnwidth]{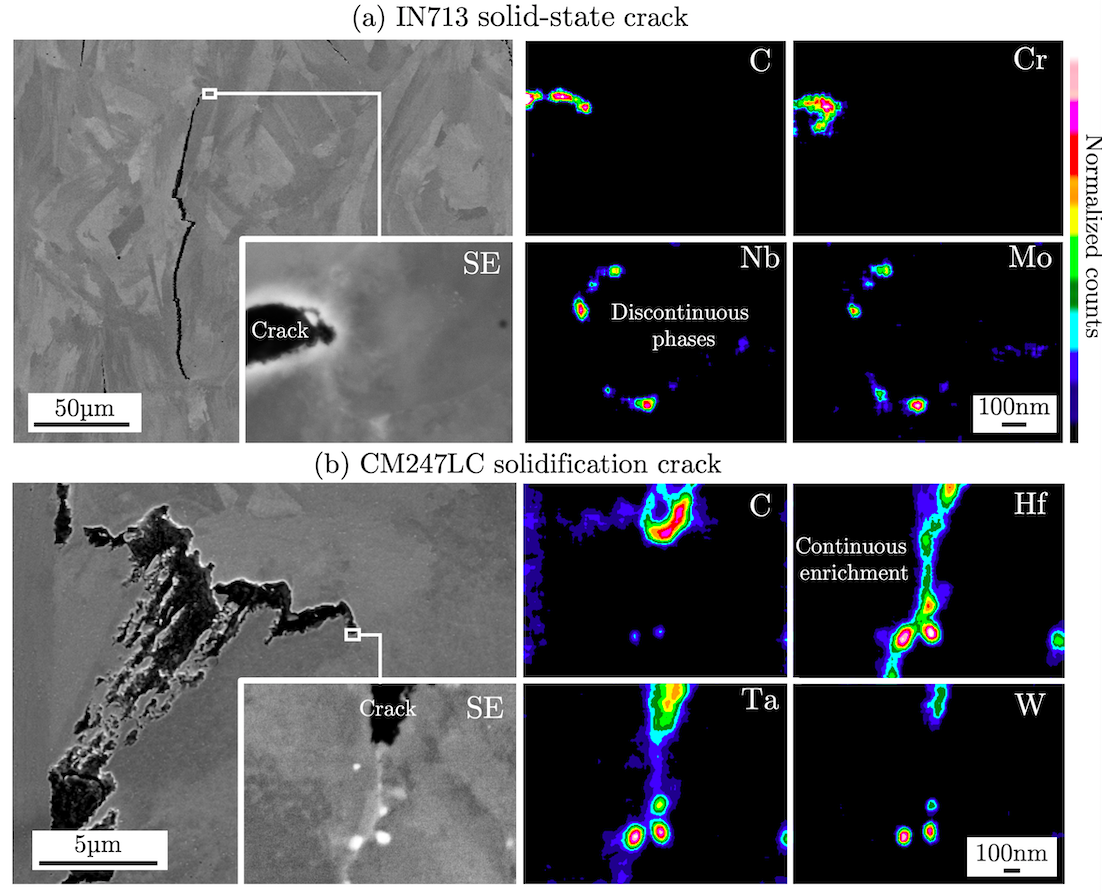}
		\caption{SEM and EDX analysis at the crack tip for a typical solid-state crack and solidification crack observed in (a) IN713 and (b) CM247LC respectively.}
		\label{figure14}
\end{figure}

In CM247LC, EDX analysis at high magnification reveals the presence of a continuous film of solute enrichment at the solidification crack tips. The enrichment of Ta and Hf separate from the carbide, shown in Figure \ref{figure14}, confirms the partitioning of these solutes to the final solidifying liquid. In IN713 -- which cracks predominantly by the solid-state mechanism -- no such continuous solute enrichment was observed, only discontinuous phases as shown in Figure \ref{figure14}. This microstructural difference between solidification and solid-state cracking confirms that CM247LC's susceptibility to solidification cracking stems from the solute enrichment of the solidifying liquid and subsequent widening of the freezing range.

\begin{landscape}
	\begin{table}[H]
		\centering
		\caption{Summary of phase transformation temperatures and carbide enrichment determined experimentally.}
		\begin{tabular}{@{}lccccll@{}}
			\toprule
			Alloy       & $T_{\mathrm{S'}}$  ($^\circ$C) & $T_{\mathrm{L}}$ ($^\circ$C) & $T_{\mathrm{L'}}$ ($^\circ$C) & $T_{\mathrm{S}}$ ($^\circ$C) & Carbide Enrichment  & Corroboration of Carbide Enrichment                                       \\ \midrule
			ABD850AM    & 1329           & 1384     & 1368    & 1264       & -     & -                                                                          \\
			ABD850AM+CB & 1276           & 1351     & 1340    & 1209      & Nb,Ta      & -                                                                          \\
			ABD900AM    & 1270           & 1363     & 1347    & 1233       & Nb, Ta & -                                                                          \\
			AM-Dev   & 1279           & 1366     & 1352    & 1229      & Nb,Ta & -                                                                          \\
			CM247LC     & 1258           & 1375     & 1366    & 1266       & Ta,Hf,W     & \cite{divya2016microstructure, wang2017microstructure}  \\
			ExpAM       & 1309           & 1361     & 1344    & 1254       & Cr,Co,Mo      & -                                                                          \\
			IN625       & 1285           & 1357     & 1308    & 1234         &  Nb      & \cite{kreitcberg2017elevated, lass2017formation}         \\
			IN713       & 1240           & 1358     & 1341    & 1238      & Nb,Mo      & \cite{lachowicz2008microstructure,chamanfar2015cracking} \\
			IN718       & 1242           & 1342     & 1296    & 1182        & Nb,Mo    &\cite{sangid2018role}                                      \\
			IN738LC     & 1279           & 1351     & 1340    & 1200       & Ta,W,Nb,Mo    & \cite{xu2018initiation, zhang2019cracking}                               \\
			IN939       & 1203           & 1326     & 1319    & 1174      & Ta,Nb   & \cite{gonzalez2011microstructural}                            \\
			Waspaloy    & 1292           & 1371     & 1355    & 1265      & Mo,Ti        &  \cite{hoier2018microstructural}                                             \\ \bottomrule
		\end{tabular}
		\label{table2}
	\end{table}
\end{landscape}

\subsubsection{Differential scanning calorimetry and thermodynamic modelling}

\begin{figure}[H]
	\centering
	\includegraphics[width=0.5\columnwidth]{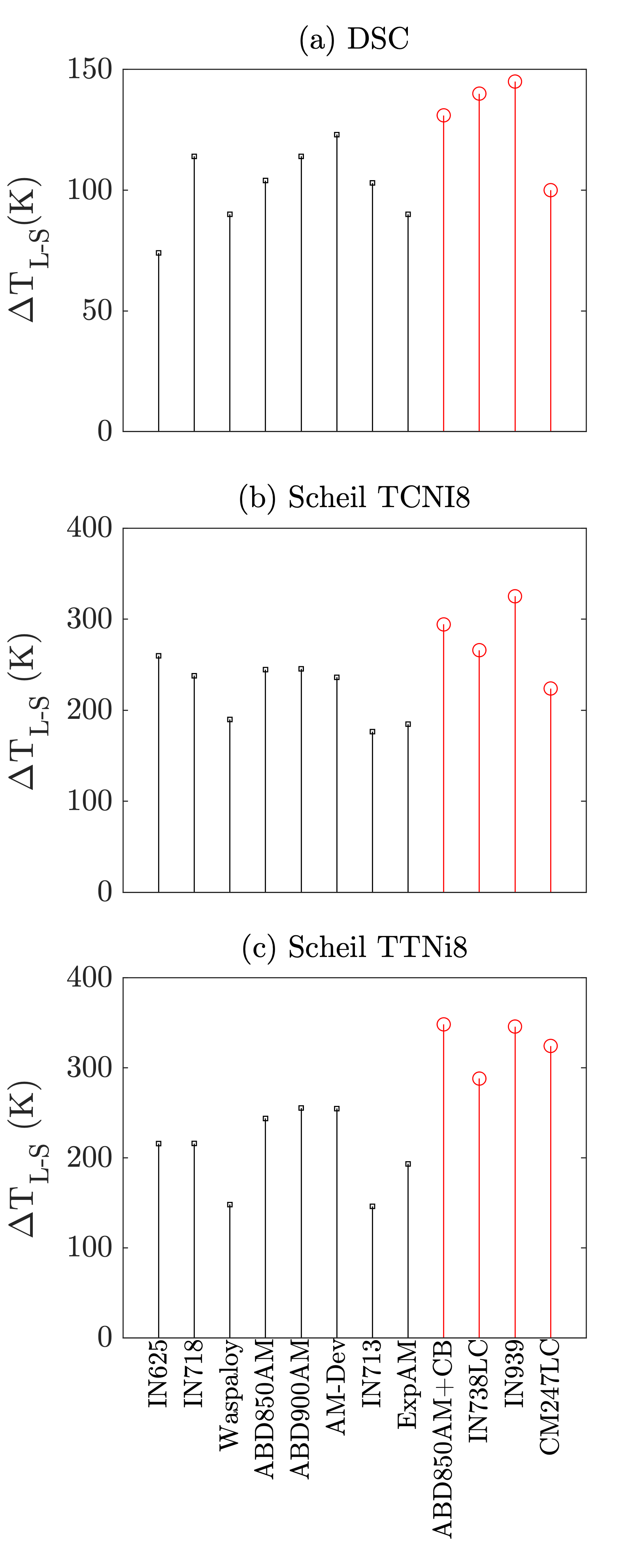}
	\caption{Summary of the magnitude of the freezing range determined (a) experimentally by DSC and by scheil model using (b) the TCNI8 and (c) the TTNi8 thermodynamic databases.}
	\label{figure15}
\end{figure}

The magnitude of the freezing range may well be a good indicator of the solidification cracking susceptibility -- the alloys prone to predominately solidification type cracks freeze over wide solidification intervals. Figure \ref{figure15} shows the magnitude of the freezing range (i) as measured by differential scanning calorimetry and (ii) and (iii) from modeling of the solidification path by thermodynamic simulation under the Scheil conditions with the TCNI8 and TTNi8 thermodynamic databases, respectively. Estimates from DSC and the TCNI8 database show the solidification cracking prone alloys ABD850AM+CB, IN738LC, and IN939 have the widest freezing range of the 12 investigated. On the other hand, TTNi8 predicts the solidification prone cracking alloys to have the 4 widest freezing ranges. 

\begin{figure}[H]
		\centering
		\includegraphics[width=1\columnwidth]{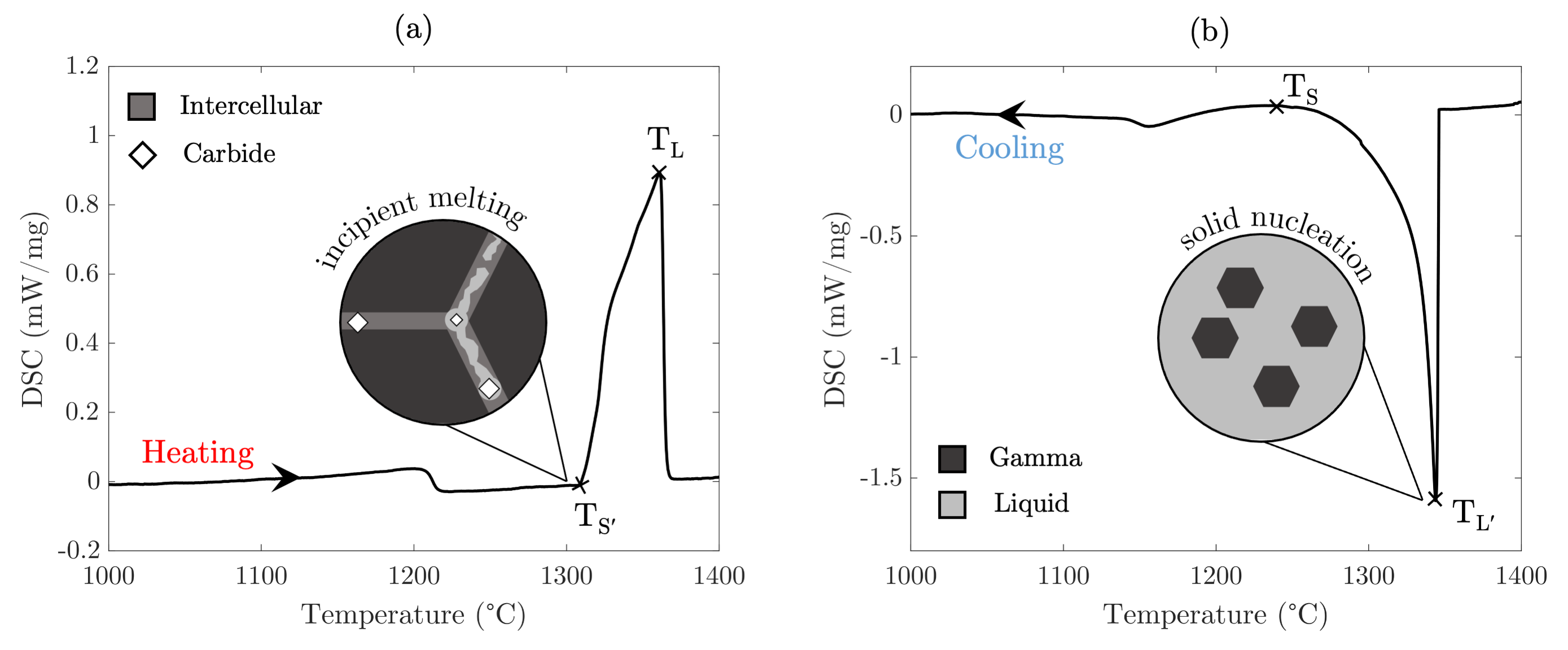}
		\caption{DSC signal from ExpAM and the respective phase transformations determined by interpretation of (a) the heating cycle and (b) the cooling cycle.}
		\label{figure16}
\end{figure}

The DSC signals and labeled phase transformation temperatures of ExpAM are shown in Figure \ref{figure16}. The simulations predict significantly lower solidus temperatures than the DSC tests and the full suppression of $\gamma^\prime$ precipitation even in the alloys with highest $\gamma^\prime$ former content; this suppression is confirmed by characterization in Figure \ref{figure1}. Liquidus and solidus temperatures determined by DSC during heating and cooling are summarized in Table \ref{table2}. One needs to bear in mind of course that it is not possible to replicate the extreme heating/cooling rates of the AM process in the DSC. 

\section{Analysis and Discussion}

The results presented above are insightful and confirm unambiguously that a strong inter-dependence exists between processability and alloy composition. Nevertheless, further understanding can be gained via more detailed analysis. We consider each of the solid-state and solidification cracking phenomena in turn.  

\subsection{Modelling of stress development during cooling leading to solid-state cracking}

\begin{figure}[H]
		\centering
		\includegraphics[width=1\columnwidth]{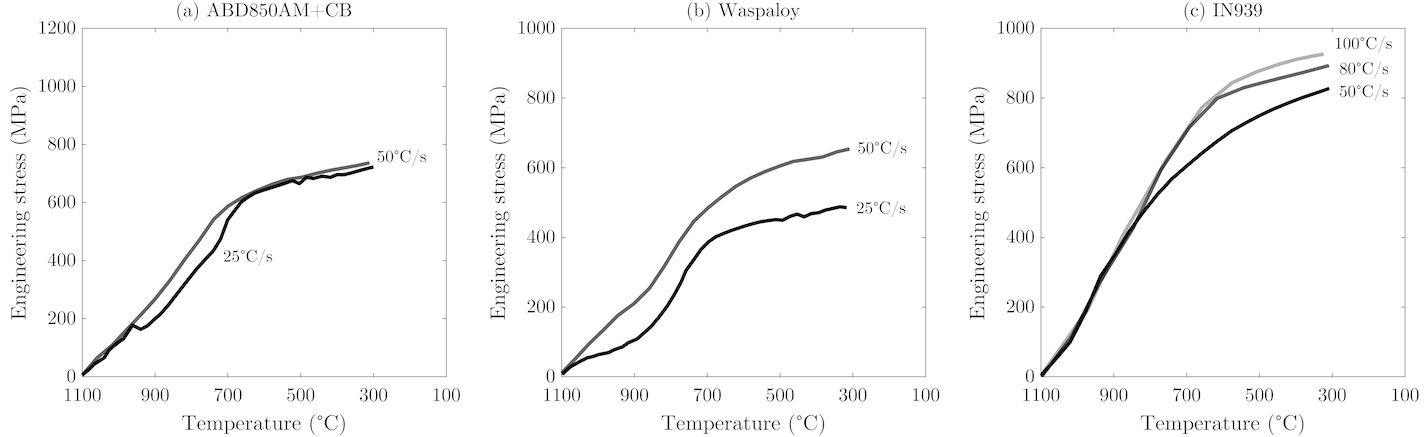}
		\caption{The stress evolution during cooling at different rates in (a) ABD850AM+CB, (b) Waspaloy, and (c) IN939.}
		\label{figure17}
	\end{figure}

\begin{figure}[H]
		\centering
		\includegraphics[width=1\columnwidth]{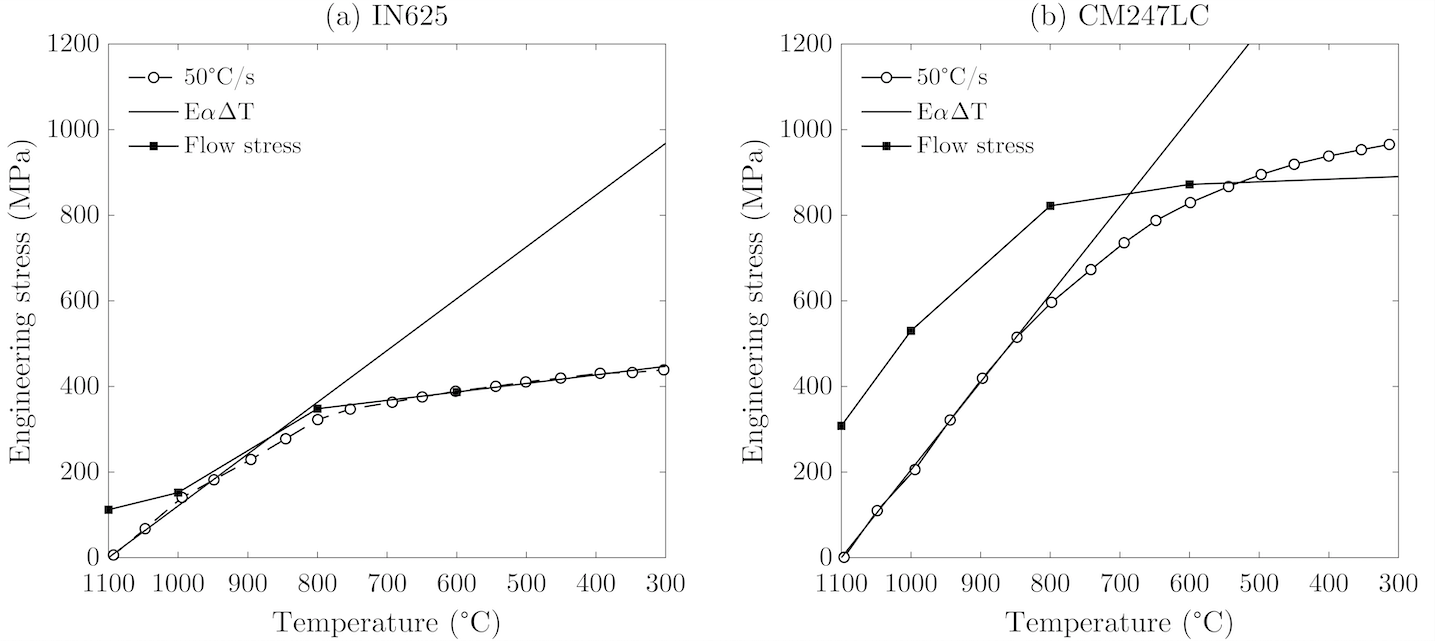}
		\caption{The stress evolution during cooling at 50$\,$$^\circ$C/s superimposed with the temperature dependent flow stress and linear thermal-elastic stress development for (a) IN625 and (b) CM247LC.}
		\label{figure18}
\end{figure}

\begin{figure}[H]
		\centering
		\includegraphics[width=1\columnwidth]{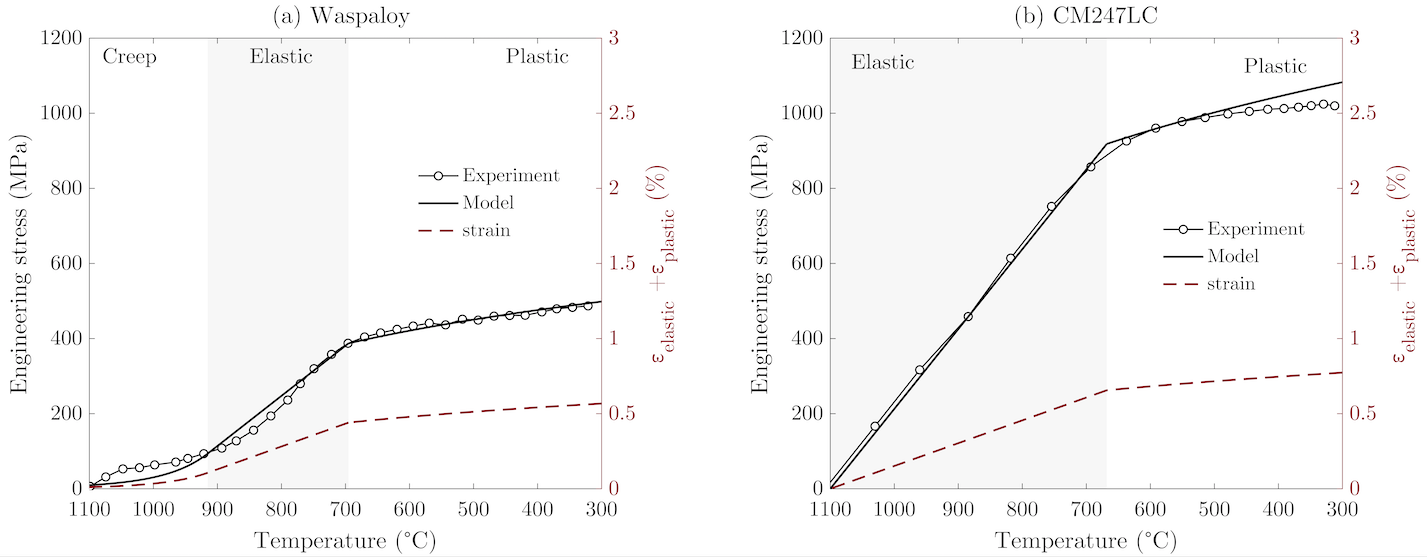}
		\caption{Comparison of the experimentally measured stress during cooling and the fitted thermal-elastic-plastic-viscoplastic model for stress (left axis) and strain (right axis) development of (a) Waspaloy and (b) CM247LC.}
		\label{figure19}
\end{figure}

Figure~\ref{figure12} confirms that the stress experienced during cooling in the constrained-bar test depends upon alloy composition. But various factors will contribute to this to varying extents -- the composition-dependence of elastic modulus and thermal expansion coefficient of Figure~\ref{figure3}, but also a possible stress relaxation effect -- and it is helpful to deconvolute these in order to determine which is the more important. Further experimentation has indicated that cooling rate does indeed influence somewhat the build up of stress, see Figure~\ref{figure17} which includes data for ABD850AM+CB, Waspaloy and IN939. Clearly, slower cooling provides more time for stress relaxation. Particularly for the stronger IN939 alloy, the initial rate of change of stress with temperature shows no dependence upon cooling rate, implying the need for interpretation based upon thermal-elastic-plastic behavior with no creep relaxation playing any role in that initial regime. Figure~\ref{figure18} provides further insight into what is happening in these tests -- consider the case of the least strong and most strong alloys considered here, IN625 and CM247LC respectfully. The variation of the measured stress with temperature for the two cases matches closely the product of elastic modulus,  thermal expansion efficient and temperature difference $E\alpha\Delta T$ for several hundred degrees below the initial temperature of 1100$\,$$^\circ$C. Particularly for the stronger alloy CM247LC, the product $E\alpha\Delta T$ lies below the measured uniaxial flow stress initially but surpasses it in the ductility-dip regime. For the weaker alloy IN625, the measured stress in the constrained bar test determined at low temperature matches well the measured flow stress consistent
with only a weak work-hardening effect. 
Modeling of the constrained bar test puts the thermal-mechanics of this situation on a quantitative basis. The
net rate of straining is zero; therefore the rates of elastic, plastic, creep and thermal straining 
$\dot \varepsilon_{\mathrm{elastic}}$, $\dot \varepsilon_{\mathrm{plastic}}$, $\dot \varepsilon_{\mathrm{creep}}$ and 
$\dot \varepsilon_{\mathrm{thermal}}$ must satisfy
\begin{equation}
	\dot \varepsilon_{\mathrm{elastic}} + 	\dot \varepsilon_{\mathrm{plastic}} + 	\dot \varepsilon_{\mathrm{creep}}+ 	\dot \varepsilon_{\mathrm{thermal}}= 0
		\label{eq2}
\end{equation}
In this expression, $\dot \varepsilon_{\mathrm{thermal}} = \alpha \dot T$ is determined by the cooling rate, we find that as the specimen is cooled one of the mechanical strain-rates ($\dot \varepsilon_{\mathrm{elastic}}$, $\dot \varepsilon_{\mathrm{plastic}}$ or $\dot \varepsilon_{\mathrm{creep}}$) dominates, with the range of dominance of each mechanism and therefore the detailed response depending on cooling rate and the mechanical properties of the material. Initially, the stress $\sigma$ equals zero. As the sample is cooled the response is elastic and 
\begin{equation}
\dot \varepsilon_{\mathrm{elastic}} = \frac{\dot \sigma}{E} = - \alpha \dot T
	\label{eq3}
\end{equation}
Giving, $\sigma = \sigma_{\mathrm{elastic}} = E \alpha \Delta T$, as shown in Figure~\ref{figure18} for IN625 and CM247LC. as the stress increases eventually creep takes over as the dominant mechanism. For a material that creeps according to Norton's Law we find
\begin{equation}
\sigma = \sigma_{\mathrm{creep}}\{T\} = \left(-\frac{\alpha  \dot T   }{A} \right)^{1/n} \mathrm{exp} \left\{ \frac{Q}{n R T} \right\}
\label{eq4}
\end{equation}
where $R$ is the ideal gas constant and the parameters approximated by fitting are the creep exponent $n$ (4.8 for both Waspaloy and CM247LC), the activation energy for creep $Q$ (800$\,$kJ/mol for Waspaloy and 1000$\,$kJ/mol for CM247LC) and the pre-exponential constant in Norton's Law $A$ ($6.1\times10^{-7}$$\,$MPa$^{-n}$$\,$s$^{-1}$ for both alloys). The actual stress is the smallest of those given by the elastic response and equation (\ref{eq4}). The initial response is elastic, but as the sample is cooled the stress increases, creep takes over as the dominant mechanism and the stress is given by equation (\ref{eq4}). For Waspaloy, which has a low creep strength at elevated temperatures, the initial elastic range is very short and creep becomes the dominant process at low stresses, as shown in Figure~\ref{figure19}a, As the temperature is decreased further the creep strength and therefore the stress given by equation (\ref{eq5}) increase. In the case of Waspaloy, we find that eventually elastic deformation dominates again. This transition occurs when
\begin{equation}
\dot \sigma_{\mathrm{creep}} =  \left( \frac{-Q}{n R {T}^2} \right) \left(-\frac{\alpha  \dot T   }{A} \right)^{1/n} \mathrm{exp} \left\{ \frac{Q}{n R T} \right\} = \dot \sigma_{\mathrm{elastic}} = E \alpha \dot T
	\label{eq5}
\end{equation}
Substituting (\ref{eq4}) into (\ref{eq5}) gives a relationship between the stress and temperature at the creep-elastic transition
\begin{equation}
\sigma_{\mathrm{ce}} = \frac{E \alpha n R {T_{\mathrm{ce}}}^2}{Q}
	\label{eq6}
\end{equation}
which we have designated using the subscript ce. The actual stress and temperature a the transition is given by the intersection of equations (\ref{eq4}) and (\ref{eq6}) on a plot of $\sigma$ against $T$. When $T < T_{\mathrm{ce}}$ the response is elastic the stress is given by
\begin{equation}
	 E \alpha (T_{\mathrm{ce}}-T) + \sigma_{\mathrm{ce}}
	 		\label{eq7}
\end{equation}
until the onset of plastic deformation.
For quantitative calculations one needs a plasticity law, taken here to be the Ramberg-Osgood relationship with 
$\sigma_{\mathrm{y}}$ the material's yield stress and $m$ a material-dependent constant, fitted to be 3.4 and 3.6 for Waspaloy and CM247LC respectively, consistent with
\begin{equation}
	\dot \varepsilon_{\mathrm{plastic}} = m \frac{\dot \sigma}{E} \left( \frac{\sigma}{\sigma_{\mathrm{y}}} \right)^{m-1}
		\label{eq8}
\end{equation}
where for simplicity we have assumed that $\sigma_{\mathrm{y}}$ does not vary with temperature. The transition to plastic deformation occurs when $\sigma$ given by equation (\ref{eq4}) and (\ref{eq7}) equals $\sigma_{\mathrm{y}}$. We designate the temperature at which this occurs as $T_{\mathrm{y}}$. It follows that subsequent response is given by
\begin{equation}
	\alpha(T_{\mathrm{y}}-T) = \frac{\sigma_{\mathrm{y}}}{E}\left[ \left( \frac{\sigma\{T\}}{\sigma_{\mathrm{y}}} \right) ^{m}   - 1 \right]
		\label{eq9}
\end{equation}
and we can describe the stress as follows for $T < T_{\mathrm{y}}$
\begin{equation}
 \sigma_{\mathrm{y}} \left( 1 + \alpha\frac{E}{ \sigma_{\mathrm{y}}} (T_{\mathrm{y}}-T) \right)^{1/m}
	\label{eq10}
\end{equation}
To summarize, the stress development during cooling can be represented by the equation set:
\begin{equation}
	\sigma\{T\} =
	\begin{cases}
		E \alpha \Delta T & \text{if $T > T_{\mathrm{ce}}$ 	$\cap$ $\sigma_{\mathrm{elastic}} < \sigma_{\mathrm{creep}}$}\\
		
		\left(-\frac{\alpha  \dot T   }{A} \right)^{1/n} \mathrm{exp} \left\{ \frac{Q}{n R T} \right\} & \text{if $T > T_{\mathrm{ce}} \cap \sigma_{\mathrm{creep}} < \sigma_{\mathrm{elastic}}$} \\
		
		E \alpha (T_{\mathrm{ce}}-T) + \sigma_{\mathrm{creep}}\{T_{\mathrm{ce}}\} &\text{if $T_{\mathrm{ce}} > T > T_{\mathrm{y}}$} \\	
		
		\sigma_{\mathrm{y}} \left( 1 + \alpha\frac{E}{ \sigma_{\mathrm{y}}} (T_{\mathrm{y}}-T) \right)^{1/m} & \text{if $T < T_{\mathrm{y}}$}
	\end{cases}
	\label{eq11}
\end{equation}

It has been found that the above formulation can largely rationalize the temperature-dependence of the build of stress with decreasing temperature, see Figure~\ref{figure19} for Waspaloy and CM247LC. All the above regimes of behavior are not exhibited by all materials. In the case of Waspaloy, the initial elastic response is small and creep dominates for the slow cooling rate considered in Figure~\ref{figure19} until approximately 900$\,$$^\circ$C when the behavior is elastic; and it becomes thermal-plastic at $\sim$700$\,$$^\circ$C. For CM247LC which displays much greater creep resistance, the deformation is initially linear thermal-elastic over a wide temperature regime until the onset of rate-insensitive plasticity again at $\sim$700$\,$$^\circ$C. In each case the temperature range over which creep occurs is small and therefore limited creep deformation occurs during cooldown. In order to compare the plastic strain accumulation to the determined composition-dependent ductility exhaustion limit, a strain-rate independent first order approximation can be made of the plastic strain as follows
\begin{equation}
	\varepsilon_{\mathrm{plastic}}\{T\} \approx E\alpha \Delta T- \frac{\sigma\{T\}}{E}
		\label{eq12}
\end{equation}
Our calculations indicate a critical minimum ductility in the range 0.5 to 1\% is likely to be needed for the avoidance of ductility-dip cracking in the critical temperature range of 700$\,$$^\circ$C to 900$\,$$^\circ$C. These findings rationalize the poor resistance of the likes of IN713 and ExpAM to solid-state cracking.

To summarize, the analysis presented in this section allows a number of critical points to be confirmed. First, the balance of evidence suggests that solid-state cracking -- in susceptible alloys -- will occur in the ductility-dip regime due to the brittleness which is prevalent there. Second, since the cooling rates of the AM process are extremely rapid \cite{li2017efficient, masoomi2017laser, panwisawas2020additive, promoppatum2017comprehensive}, the development of stress will be thermal-elastic-plastic in origin, such that for the highest strength alloys there will be little influence of stress relaxation on the composition-dependence of cracking susceptibility. Third, nonetheless the alloy composition is important: the greater the extent of alloying, the greater the stress build up to drive cracking and the greater the risk of the yield stress being exceeded in the ductility-dip regime. Thus, the cracking risk is exacerbated by greater alloy strength but for any given strength, the risk is ameliorated if the low-temperature ductility is greater. These two factors are strongly correlated via the strength/ductility relationship. Finally, the critical ductility needed to avoid solid-state cracking appears to be in the range 0.5 to 1\% in the ductility dip regime. Obviously, the above applies only to solid-state cracking and ignores the solidification cracking effect which is now considered in the following section.

\subsection{Analysis of solidification cracking susceptibility}

So far, the occurrence of solidification cracking has been correlated with the magnitude of the freezing range and presence of continuous films of solute enrichment at crack tips. But a number of composition-dependent solidification cracking criteria should be considered, for example those due to Clyne \& Davies (CD) \cite{clyne1981influence}, Rappaz, Drezet, \& Gremaud (RDG) \cite{rappaz1999new} and Kou \cite{kou2015criterion}. In each case, a parameter $\Phi$ can be identified, which is predicted to scale with cracking susceptibility. The CD approach for instance considers the dimensionless ratio of the time during solidification during which the alloy is vulnerable to cracking $\Delta t_{\mathrm{vulnerable}}$ to the time during which it can relieve the accumulated stress $\Delta t_{\mathrm{stress \: relief}}$. Assuming a constant cooling rate this ratio is proportional to each respective temperature range, consistent with
\begin{equation}
	\Phi_{\mathrm{CD}}\propto \frac{\Delta T_{\mathrm{vulnerable}}}{\Delta T_{\mathrm{stress \: relief}}} = 
	\frac{ T_{f_{\mathrm{s}}=0.99}-T_{f_{\mathrm{s}}=0.9}}{T_{f_{\mathrm{s}}=0.9}-T_{f_{\mathrm{s}}=0.4}}
	\label{eq:13}
\end{equation}
The CD criterion can be criticized on account of its rather arbitrary definition of the range during which the material is vulnerable; however, keeping this constant for the 12 alloys allows for comparison. The alternative RDG criterion is based on a balance of pressure, such that the formation of hot tears depends upon the maximum strain rate $\dot{\varepsilon}_{\mathrm{max}}$ before a cavitation pressure $\Delta p_\mathrm{c}$ results in the formation of a void. Both the mechanical contribution and the ability for liquid to back fill are dictated by integrals of the solid fraction vs temperature $f_{\mathrm{s}}(T)$, $A$, $B$, and $C$ \cite{rappaz1999new} consistent with
\begin{equation}
	\Phi_{\mathrm{RDG}} \propto \frac{1}{\dot \varepsilon_{\mathrm{max}}} = \frac{(1+\beta)B}{G} 
	\bigg[  \frac{\Delta p_{\mathrm{c}} G \lambda_2^2 }{180 \mu} -  v_{\mathrm{T}} \beta A   \bigg]^{-1}
	\label{eq:14}
\end{equation}
where
\begin{equation}
	A = \int_{T_{\mathrm{cg}}}^{T_{\mathrm{L}}} \frac{f_{\mathrm{s}}\{T\}^2}{(1-f_{\mathrm{s}}\{T\})^2} \mathrm{d}T
		\label{eq15}
\end{equation}
and
\begin{equation}
	B = \int_{T_{\mathrm{cg}}}^{T_{\mathrm{L}}} \frac{C}{(1-f_{\mathrm{s}}\{T\})^3} \mathrm{d}T
		\label{eq16}
\end{equation}
where
\begin{equation}
	C = f_{\mathrm{s}}\{T\}^2 \int_{T_{\mathrm{\mathrm{cg}}}}^{T_{\mathrm{L}}} f_{\mathrm{s}}\{T\} \mathrm{d}T
		\label{eq17}
\end{equation}
For the calculations carried out here, the viscosity $\mu$, the thermal gradient $G$, the velocity of the isotherms $v_{\mathrm{T}}$, and the shrinkage factor $\beta$ have been taken from \cite{rappaz1999new}, the secondary arm spacing $\lambda_2$ from \cite{tang2021alloys}. The grain coalescence temperature $T_{\mathrm{cg}}$ is taken to correspond to $T_{f_{s}=0.99}$. The hot cracking susceptibility is proportional to $\dot{\varepsilon}^{-1}_{\mathrm{max}}$ and values are normalized between zero and unity. The RDG criterion suffers from a similar criticism to the CD criterion, as the grain coalescence temperature -- critical to evaluating the integrals $A$, $B$ and $C$ -- is not well understood. Furthermore, approximating liquid filling by the flow of fluid through a packed bed may not be the most appropriate for the cellular AM microstructures. Finally, the Kou solidification cracking index is formulated around the balance of increasing spatial dimensions associated with solid formation and volumetric liquid flow within a differential control volume between two growing grains \cite{kou2015criterion}. Here, the cracking susceptibility is taken to be proportional to the maximal value of the ratio of change in temperature $\mathrm{d}T$ relative to the change in square root in the solid fraction $\mathrm{d}(\sqrt{f_{\mathrm{s}}})$ near the final stages of solidification, consistent with
\begin{equation}
	\Phi_{\mathrm{Kou}} \propto \bigg|\frac{\mathrm{d}T}{\mathrm{d}(\sqrt{f_{\mathrm{s}}})} \bigg|_{f_{\mathrm{s}} \rightarrow 1}
	\label{eq19}
\end{equation}

\begin{figure}[H]
	\centering
	\includegraphics[width=0.9\columnwidth]{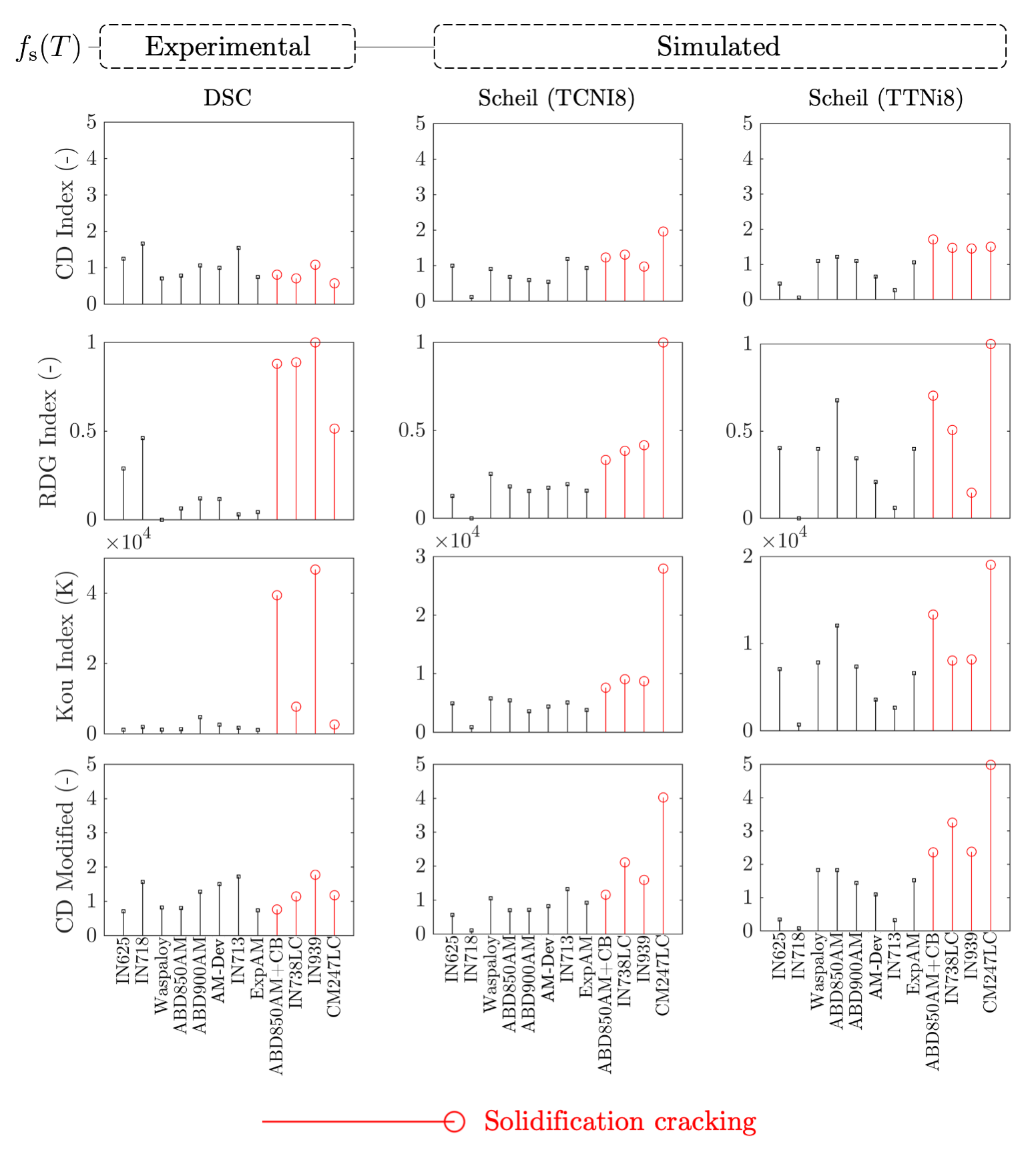}
	\caption{Solidification cracking susceptibility as predicted by the Clyne-Davies criterion, the Rappaz-Drezet-Gremaud criterion, the Kou criterion, and a modified Clyne-Davies criterion,  each determined as a function of the solidification path determined experimentally and via scheil with the TCNI8 and TTNi8 databases.}
	\label{figure20}
\end{figure}

Figure \ref{figure20} summarizes the predictions of the different criteria considered. One can see that, with the exception of the CD index evaluated with the experimentally-evaluated DSC traces (which probably gives an underestimate of the solidification range) -- the indices are in line with the findings of our experimentally-determined solidification cracking susceptibility. However, no single index predicts correctly the composition-dependence of solidification cracking for all methods of determining $f_{\mathrm{s}}\{T\}$. Moreover, no one method of estimating $f_{\mathrm{s}}\{T\}$ correctly predicts solidification cracking across all the indices. The strength of applying these legacy criteria to multi-component Ni-superalloys is their sensitivity to the final stages of solidification whereby the final liquid is solute enriched and stable at lower temperatures. This in turn presents a problem, as $f_{\mathrm{s}}\{T\}$ is discontinuous when a new phase becomes stable, making the exact numerical procedure by which the Kou criterion is evaluated highly sensitive to these discontinuities. 

One can argue that a modified criterion is needed to take account of contributions of mechanical and stress relaxation properties. A very detailed analysis is beyond the scope of this paper, but approximate calculations can be made as follows. Assume the contribution of stress relaxation is more pronounced at near solidus temperatures as is suggested by the strong temperature dependence of stress relaxation of Figure \ref{figure11}. A modified criterion could consider the solidification cracking susceptibility to be proportional to the ratio of strain accumulated during the vulnerable regime $\alpha \Delta T_{\mathrm{vulnerable}}$ to the strain relieved in the stress relief regime. We apply this as shown in Equation \ref{eq20} using experimentally determined values of $E$, and $\alpha$, and the $\mathrm{d}\sigma/\mathrm{d}t$ derived from the secant line of the stress relaxation behavior from 0 to 30$\,$s at 1000$\,$$^{\circ}$C, according to:
\begin{equation}
	\Phi_{\mathrm{CD  \: modified}}  \propto \frac{\varepsilon_{\mathrm{vulnerable}}}{\varepsilon_{\mathrm{stress \: relief}}} =  \frac{\alpha \Delta T_{\mathrm{vulnerable}}}{\left( \frac{\dot \sigma}{E} \right) \frac{\Delta T_{\mathrm{stress \: relief}}}{\partial T/ \partial t}}
	\label{eq20}
\end{equation}
This is in essence a mechanically-weighted CD index. The modified index increases the relative predicted susceptibility of the solidification cracking alloys IN738LC, IN939, and CM247LC appreciably, see Figure \ref{figure20}, by accounting for thermal expansion, stiffness and the capacity of the material to stress relax.

\section{Final considerations}

\begin{figure}[H]
		\centering
		\includegraphics[width=1\columnwidth]{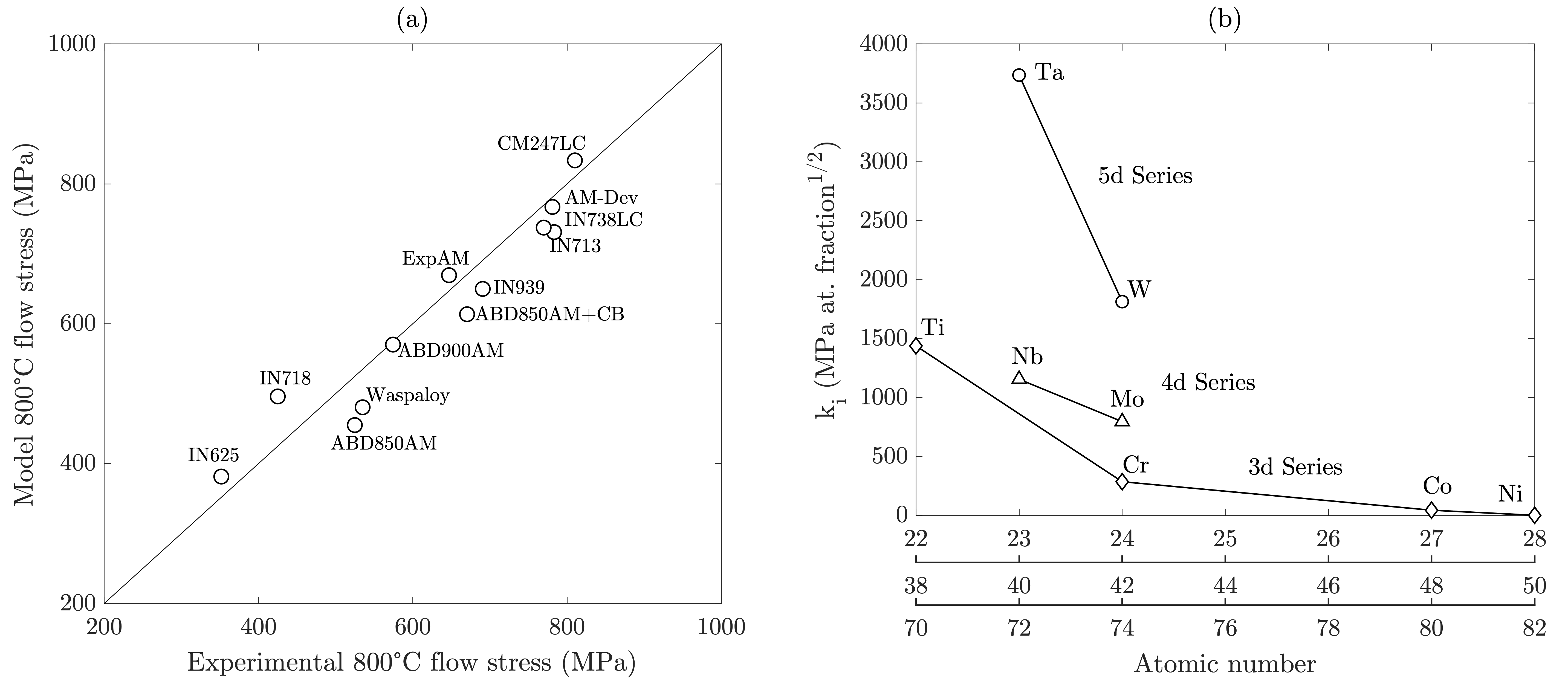}
		\caption{(a) Comparison of the experimentally determined flow stress and the solid solution strengthening model at 800$\,$$^\circ$C and  (b) strengthening coefficients as a function of atomic number for solutes in the 3d, 4d, and 5d groups.}
		\label{figure21}
\end{figure}

An intriguing aspect of the as-processed microstructures produced in the present work is the absence of the $\gamma^\prime$ phase due to the suppression of the $\gamma\rightarrow\gamma+\gamma^\prime$ reaction caused by the rapid cooling rates achieved. It follows that the varying strength levels measured arise from different degrees of solid-solution hardening, since the same processing parameters yield similar in-plane grain size distributions and thus approximately similar degrees of sub-structure hardening via the Hall-Petch effect. Thus one has a very supersaturated solution of hardening elements present in significant concentrations in the matrix $\gamma$ phase. For this situation, there is a complete paucity of information in the literature for the relative contributions of the elements to the strengthening levels achieved in this situation. Although measurements have been made on just 12 different alloys in this work, it has proven possible to come to interesting findings in the following way. Assume that the contributions to the strengthening of each element $i$ via a coefficient $k_i$ can be summed according to a simple model consistent with \cite{gypen1977multi}
\begin{equation}
	\sigma_{\mathrm{solid \: solution}} = \bigg( \sum_{i}  k_{i} ^ \frac{1}{n} x_i \bigg) ^ n
		\label{eq21}
\end{equation}
where $x_i$ the concentration of solute $i$, $k_i$ is the respective strengthening constant and the value of $n$ is taken as 2/3 \cite{gypen1977multi}. A multi-regression fitting procedure then allows an estimate of the values of $k_i$, see Table~\ref{table3}. The values obtained are compared with those determined by Roth et al \cite{roth1997modeling} on much more dilute alloys, via a trawling of the literature. Similar trends are observed, which lends confidence to our strength measurements. When the coefficients $k_i$ are plotted as a function
of the location of element $i$ in the d-block of transition metals, one sees that the elements from the far-west of the d-block confer a greater degree of hardening, on an atomic percent basis. Moreover, it would appear that the extent of hardening for each column in the d-block is consistent with $\rm 5d > 4d > 3d$, see Figure \ref{figure21}. Our data are sparse, but our findings are worthy of future investigations along these lines.

\begin{table}[H]
	\centering
	\caption{Strengthening coefficients for solution strengthening in nickel for AM at 800$\,$$^\circ$C in comparison to \cite{roth1997modeling} (MPa at Fraction$^{1/2}$).}
	\begin{tabular}{@{}ccc@{}}
		\toprule
		Alloying   Element & $k_{\mathrm{i}}^{\mathrm{AM}}$    & $k_{\mathrm{i}}$ \cite{roth1997modeling} \\ \midrule
		Al                 & 1787 & 225        \\
		Co                 & 43   & 39         \\
		Cr                 & 285  & 337        \\
		Fe                 & -    & 153        \\
		Mo                 & 795  & 1015       \\
		Nb                 & 1155 & 1183       \\
		Ta                 & 3736 & 1191       \\
		Ti                 & 1438 & 775        \\
		W                  & 1813 & 977        \\
		C                  & 5355 & 1061       \\
		Hf                 & -  & 1401       \\ \bottomrule
	\end{tabular}
	\label{table3}
\end{table}

\begin{figure}[H]
		\centering	\includegraphics[width=1\columnwidth]{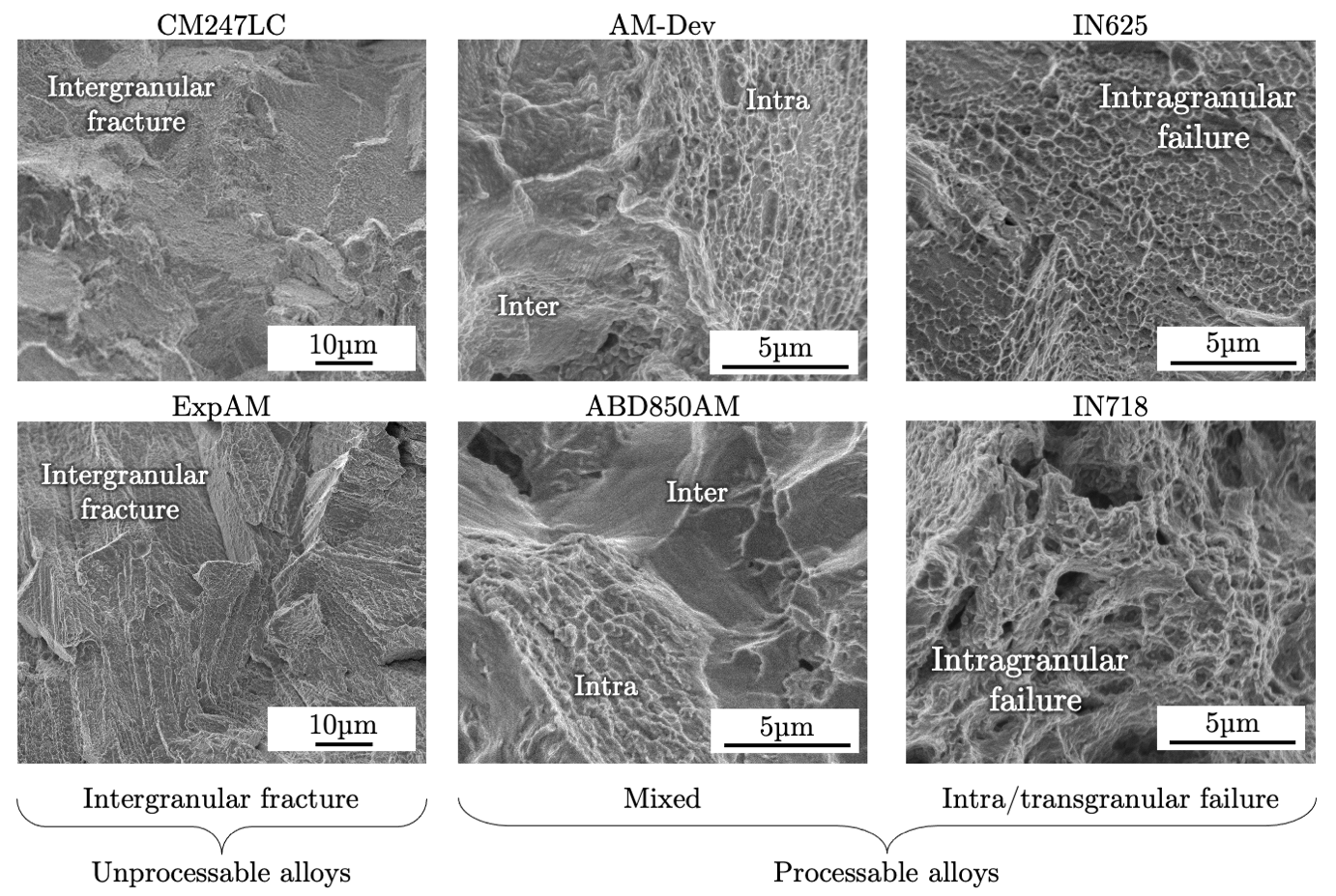}
		\caption{Fracture surfaces of CM247C, AM-Dev, and IN625 after tensile tests at 800$\,$$^\circ$C illustrating the transition from intra/transgranular to intergranular fracture.}
		\label{figure22}
\end{figure}

\begin{figure}[H]
		\centering
		\includegraphics[width=1\columnwidth]{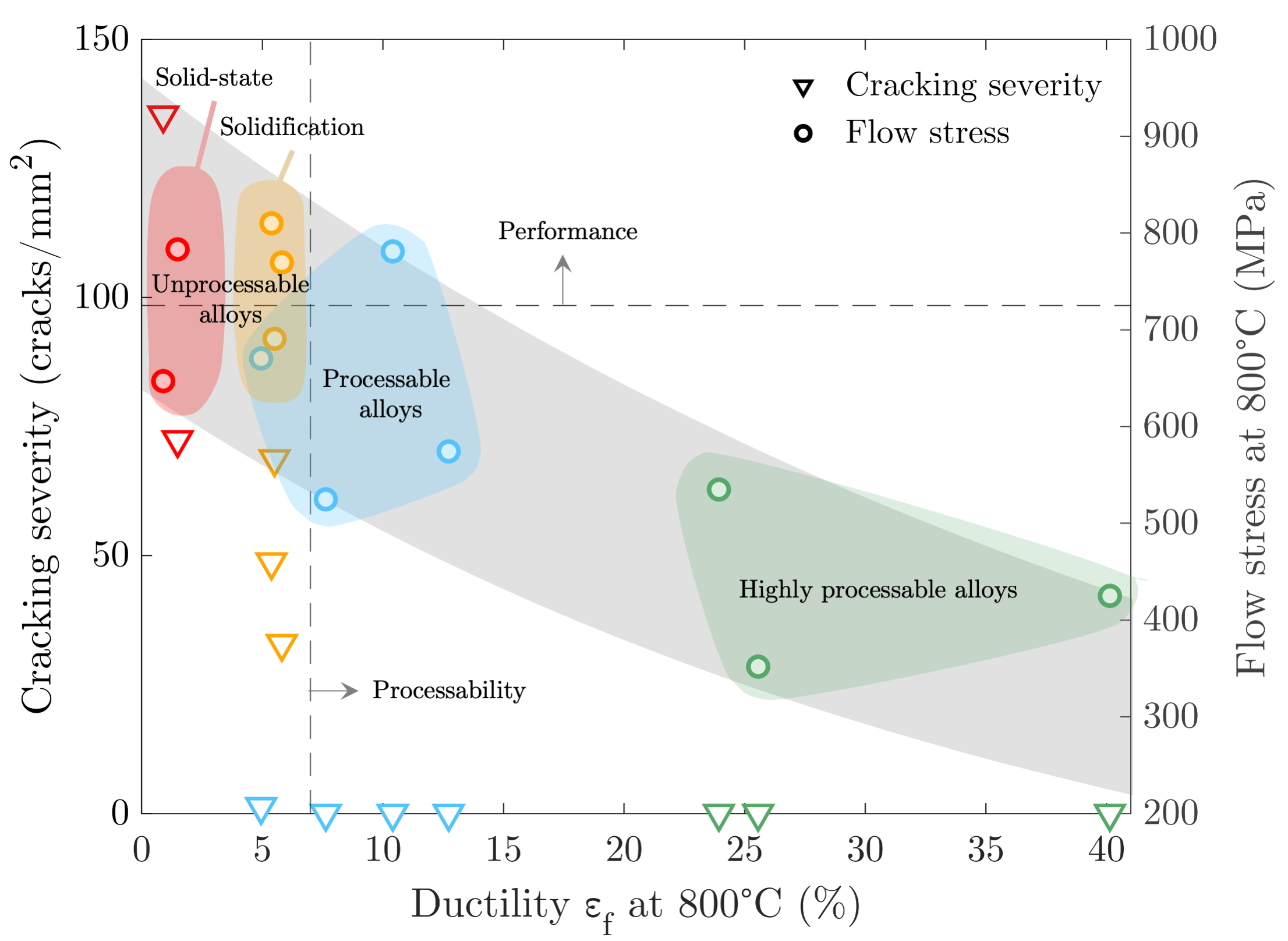}
		\caption{Processing map demonstrating four classes of alloys based upon measurements of their cracking severity (left axis) and flow stress at 800$\,$$^\circ$C (right axis) as a function of their the ductility at 800$\,$$^\circ$C.}
		\label{figure23}
\end{figure}

Despite previous work having suggested that improving the solid solution strengthening can improve processability \cite{harrison2015reduction}, the research presented here indicates that it is the exhaustion of ductility -- not strength -- which must be conferred for crack-free printing if the best strength levels are to be approached. This is summarized succinctly in Figure~\ref{figure23}. Plotting the alloys' ductility vs cracking severity indicates alloys become unprocessable when their ductility is less than 7\%, though it should be noted that it would appear to be possible to achieve strength levels comparable to those of CM247LC without embrittlement. In general, a strength/ductility trade-off is apparent as the macroscopic elongation in the ductility-loss regime ($\sim$800$\,$$^\circ$C) is governed by the balance of plasticity of grain interiors and the cohesion of grain boundaries. For very strong alloys -- such as CM247LC and ExpAM -- the failure mechanism is purely intergranular -- see Figure~\ref{figure22} -- implying grain boundaries are the ductility-limiting feature of the unprocessable alloys. Whereas for alloys such as IN718 and IN625, the fracture surfaces show only transgranular features such as those due to micro-void coalescence -- suggesting grain boundaries are sufficiently cohesive, making the grain interior the ductility-limiting feature. The fracture surfaces of the moderate ductility ABD alloys such as ABD850AM and 1005AM are somewhere in between: they exhibit features characteristic of both intergranular and transgranular fracture, see Figure~\ref{figure22}. This work finds that despite the presence of such carbides in the processable and ductile Nb/Mo-rich carbide forming alloys IN625, IN718 and the ABD alloys, their occurrence in the un-processable IN713 alloy suggests that carbide enrichment alone does not guarantee grain boundary properties. It seems likely that the ratio of grain boundary strength to grain strength -- in this case controlled by the contribution of elements in solid solution -- dictates the extent to which fracture is inter- or intragranular. These need to be balanced to maximize performance whilst maintaining processability. Clearly, further advances in the field of grain boundary engineering are needed if further improvements in property levels are to be achieved. 


\newpage

\section{Summary and Conclusions}

In this paper, the additively manufacturability of nickel-based superalloys has been considered in detail. The laser powder bed fusion technique is used. Twelve different alloys have been studied, some of the heritage type but also some new ones designed specifically for this process. Emphasis has been placed on the influence of composition on cracking susceptibility. The following specific conclusions can be drawn from this work:

\begin{enumerate}
	
	\item A dependence of alloy composition on processability has been proven. Of the twelve alloys studied, quantitative stereology confirmed that half are susceptible to processing-induced crack defect formation under the experimental conditions employed. In particular, IN713 and ExpAM compositions were shown to be prone to solid-state cracking, whilst CM247LC, IN738LC, ABD850AM+CB and IN939 exhibited a tendency towards solidification cracking.
	
	\item Solid-state cracking arises from a significant loss of tensile ductility at 800$\,$$^\circ$C  ($<$7\%) which was prevalent in all unprocessable alloys -- as exemplified in particular by the IN713 and ExpAM alloys for which it was $<$1\%. Stereological analysis indicates that this brittleness is intergranular in form implying grain boundary weakness and exacerbated by increased $\gamma^\prime$ former content.
	
	\item A thermal-elastic-plastic-viscoplastic analysis of a 1D constrained bar has shed light on the factors exacerbating solid-state cracking. It seems likely that this effect is not strongly affected by creep-driven stress-relaxation, under the cooling rates experienced. Instead, processing is compromised by brittleness in the ductility-dip regime. Nevertheless, at any given strength level it appears to be possible to find alloys which are processable, and others which are not.  
	
	\item When solidification cracking occurs, it is in alloys that freeze over a wide temperature interval and relieve stress (by creep processes) at a slower rate. Increased C content and the presence of of Hf/Ta enriched crack tips is characteristic of the alloys which cracked by the solidification mechanism.
	
	\item Solidification cracking criteria following Clyne-Davies, Rappaz, and Kou have been considered as a means of rationalizing our findings concerning solidification cracking; these are shown to have some predictive power. However, they do not account for the composition-dependent near-solidus mechanical properties which may be important for the stress relaxation processes needed for crack-free processing. A modified Clyne-Davies model is proposed to account for this.
	
	\item The processing/performance relationships of existing and newly designed alloys have been compared. It has been shown that the strength levels of the former can be approached whilst engendering the AM processability which is apparently lacking.
	
	\item Our results suggest pathways by which the new alloys may be further improved. Since creep relaxation does not seem to play a role in the solid-state cracking phenomena, it might prove possible to design alloys of creep resistance equivalent to the very best conventionally-cast grades which are amenable to AM processing, provided that they resist solidification cracking.
	
\end{enumerate}

\section{Acknowledgments}

The financial support of this work by Alloyed Ltd. as well as the the The Natural Sciences and Engineering Research Council of Canada (NSERC) in the Chemical, Biomedical and Materials Science Engineering division award number 532410. The authors acknowledge funding from Innovate UK, under project number 104047, specifically the Materials and Manufacturing Division. 

\section{References}

\bibliographystyle{MetTransA}
\bibliography{ghoussoub2021on}

\end{document}